\documentclass[a4]{article}
\pdfoutput=1

\usepackage[dvipdfmx]{graphicx}
\usepackage{subfig}
\usepackage{mediabb}
\usepackage{amsmath}
\usepackage{amssymb}
\usepackage{xcolor}
\usepackage{url}
\usepackage{mathtools}
\usepackage{stmaryrd}
\usepackage{cite}
\usepackage[version=3]{mhchem}
\usepackage{authblk}
\usepackage{chngcntr}

\def\*#1{\boldsymbol{#1}}
\newcommand{\tw}[0]{\textwidth}

\newcommand{\argmax}{\mathop{\text{argmax}}}

\newcommand{\lw}[1]{\smash{\lower2.ex\hbox{#1}}}

\newcommand{\cbr}[1]{\left\{#1\right\}}

\newcommand{\RR}{\mathbb{R}}

\newcommand{\EE}{\mathbb{E}}

\newcommand{\cN}{{\cal N}}

\newcommand{\cS}{{\cal S}}

\newcommand{\cY}{{\cal Y}}

\title{
Computational Design of Stable and Highly Ion-conductive Materials using Multi-objective Bayesian Optimization: \\
Case Studies on Diffusion of Oxygen and Lithium 
}
\author[1,3,4]{Masayuki Karasuyama\footnote{corresponding author: karasuyama@nitech.ac.jp}}
\author[2]{Hiroki Kasugai}
\author[2,4]{Tomoyuki Tamura}
\author[5,6,4]{Kazuki Shitara}

\affil[1]{Department of Computer Science, Graduate School of Engineering, Nagoya Institute of Technology, Gokiso-cho, Showa-ku, Nagoya, 466-8555, Japan}
\affil[2]{Department of Physical Science and Engineering, Graduate School of Engineering, Nagoya Institute of Technology, Gokiso-cho, Showa-ku, Nagoya, 466-8555, Japan}
\affil[3]{JST, PRESTO, 4-1-8 Honcho, Kawaguchi, Saitama, 332-0012, Japan}
\affil[4]{Center for Materials Research by Information Integration, National Institute for Materials Science, 1-2-1 Sengen, Tsukuba, Ibaraki, 305-0047, Japan}
\affil[5]{Joining and Welding Research Institute, Osaka University, 11-1 Mihogaoka, Ibaraki city, Osaka, 567-0047, Japan}
\affil[6]{Nanostructures Research Laboratory, Japan Fine Ceramics Center, Nagoya 456-8587, Japan}

\date{}

\usepackage[margin=1in]{geometry}

\newif\ifdraft % \drafttrue 
\draftfalse

\begin{document}

\maketitle

\begin{abstract}
Ion-conducting solid electrolytes are widely used for a variety of purposes.
Therefore, designing highly ion-conductive materials is in strongly demand.
Because of advancement in computers and enhancement of computational codes, theoretical simulations have become effective tools for investigating the performance of ion-conductive materials.
However, an exhaustive search conducted by theoretical computations can be prohibitively expensive.
Further, for practical applications, both dynamic conductivity as well as static stability must be satisfied at the same time.
% not only dynamic conductivity but also static stability must be satisfied at the same time.
%
Therefore, we propose a computational framework that simultaneously optimizes dynamic conductivity and static stability; this is achieved by combining theoretical calculations and the Bayesian multi-objective optimization that is based on the Pareto hyper-volume criterion.
Our framework iteratively selects the candidate material, which maximizes the expected increase in the Pareto hyper-volume criterion; this is a standard optimality criterion of multi-objective optimization.
Through two case studies on oxygen and lithium diffusions, we show that ion-conductive materials with high dynamic conductivity and static stability can be efficiently identified by our framework.
\end{abstract}

\noindent
{\bf keywords: } Ion-conductive material, Bayesian multi-objective optimization, Pareto hyper-volume

% \tableofcontents

\clearpage

% --------------------------------------------------
\section{Introduction} \label{sec:introduction}

% Ion-conducting solid electrolytes are widely used as solid oxide fuel cells (SOFC) with proton conducting or oxide-ion conducting, next-generation all solid state lithium-ion batteries (LIB) with lithium-ion conducting, and resistive random access memory (ReRAM).
Ion-conducting solid electrolytes are widely used in a variety of devices such as solid oxide fuel cells (SOFC) with proton or oxide-ion conduction, next-generation all solid state lithium-ion batteries (LIB) with the lithium-ion conduction, and resistive random access memory (ReRAM).
%
% Ion-conducting solid electrolytes are widely used in solid oxide fuel cells (SOFC) with next-generation all-solid-state lithium-ion batteries (LIB) that exhibit proton or oxide-ion conduction, lithium-ion conduction, and a resistive random access memory (ReRAM).
%
% To improve the ion conductivity of solid electrolytes, elucidation of the diffusion mechanism of ions included in high concentration has been extensively studied from both experiments and theoretical calculations.
To improve the ion conductivity of solid electrolytes, elucidation of the diffusion mechanism of ions used in a high concentration has been extensively studied via both experiments and theoretical calculations.

% Because of advances in computers and enhancement of computational codes, theoretical simulations are possible even for dynamic characteristics requiring large computational costs.
% Because of advances 
Recent enhancement on computers and computational codes, theoretical simulations are possible even for dynamic characteristics possessing large computational costs.
%
% As a result, %diffusion path and diffusion activation energy
% diffusion paths and those activation energies have been clarified for many systems.
As a result, diffusion paths and the corresponding activation energies have been evaluated for many systems.
%
% In the future, designing high ion-conductive materials is required by optimizing constituent elements and composition ratios.
Thus, designing highly ion-conductive materials by optimizing constituent elements and composition ratios is a natural step forward.
%
% However, for practical applications, not only dynamic conductivity but also static stability must be satisfied at the same time.
However, for practical applications, both dynamic conductivity and static stability must be satisfied simultaneously.
%
% Since the search space is enormous and the trade-off relation often exists in conductivity and stability, development of an efficient search method is strongly demanded for ion-conductive materials.
Since the search space is enormous and the trade-off relation between conductivity and stability generally exists, the development of an efficient search method is necessitated for ion-conductive materials.
%
% Nowadays, materials informatics researches combining theoretical calculations and machine-learning methods have been actively performed.
Nowadays, research on materials informatics is being actively performed by combining theoretical calculations and machine-learning methods.
%
% In particular, Bayesian optimization has been shown as an effective exploration algorithm for material discovery \cite{Seko2015,Toyoura2016,Packwood2017,Yonezu2018}.
In particular, Bayesian optimization has been shown to be an effective exploration algorithm for material discovery (e.g., \cite{Seko2015,Toyoura2016,Packwood2017,Yonezu2018}).
In this study, we consider the simultaneous optimization of dynamic conductivity and static stability through Bayesian optimization for ion-conductive materials.
We propose a computational framework that combines theoretical calculations and the Bayesian multi-objective optimization with the \emph{Pareto hyper-volume} criterion.

%
% We show that our Gaussian process based statistical decision making can drastically accelerate the exploration of ion-conductive materials. 

\begin{figure}[t]
 \centering
 \includegraphics[clip,width=0.6\tw]{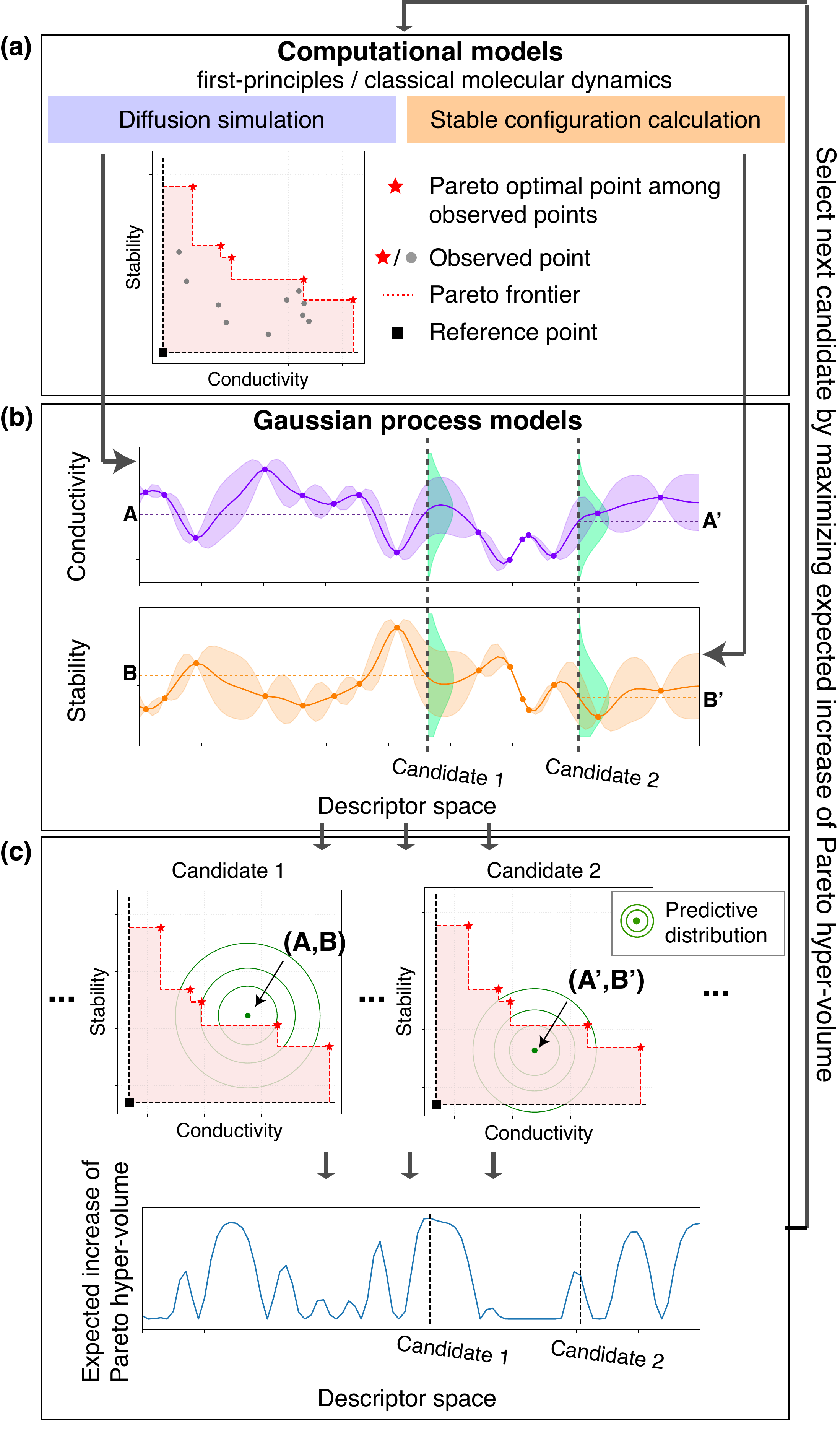}   
 \caption{
 Entire flow of our exploration framework of ion-conductive materials.
 (a) Stability and conductivity evaluation by theoretical calculations.
 The red stars and gray points are the observed points, and the red stars are the Pareto optimal points among them.
 The quality of the observed points can be evaluated by the Pareto hyper-volume that is defined as the area of the light-red region. 
 To define the volume, a reference point is defined as the lower bound of each axis.
 The dashed red line in the plot is called Pareto frontier, which is the boundary defined by the current optimal points.
 If a new point is observed across this boundary to the right and/or upper side of the plot, the Pareto frontier is augmented.
 (b) GP based modeling of conductivity and stability.
 For the observed conductivity and stability, GP is fitted in a descriptor space of candidate materials.
 (c) Evaluation of improvement of Pareto optimality.
 For new candidate points, we evaluate the predictive distributions (green distributions) of GP.
 By assuming that a new point is obtained from this distribution, we evaluate the increase of the Pareto hyper-volume as an expected value.
 In this illustration, candidate 1 achieved the maximum expected increase in the volume.
 }
 \label{fig:overview}
\end{figure}

Figure~\ref{fig:overview} shows the overview of our framework.
We first evaluate stability and conductivity by using first-principles molecular dynamics (FPMD) or classical molecular dynamics (CMD) as shown in Fig.~\ref{fig:overview}~(a).
We consider the Pareto optimality as an evaluation measure of the observed points, which is a standard optimality criterion in the multi-objective optimization literature \cite{Emmerich2006}. 
With regard to the Pareto optimal point, there does not exist any other points that can improve every objective function simultaneously (indicated by red stars in (a)).
Therefore, by identifying a set of the Pareto optimal points, we can obtain a set of materials with multiple superior properties.
Figure~\ref{fig:overview}~(b) indicates that the Gaussian process (GP) models are fitted to the calculated stability and conductivity.
In each figure of (b), the horizontal axis represents a descriptor space that is constructed from the atomic configurations and/or the atomic species of the candidate materials.
From GP, we can also obtain uncertainty of the prediction (shaded regions in (b)).
For any one of candidate materials, uncertainty is represented as a Gaussian distribution (vertical green distribution).
As shown in the figure, uncertainty becomes large around the region where observations do not exist (i.e., the region that has not been explored).
In Fig.~\ref{fig:overview}~(c), a probabilistic prediction of the next candidate point is shown in the form of the green distribution.
To estimate possible improvement by sampling a new candidate, we employ the \emph{expected increase of the Pareto hyper-volume} (EIPV) \cite{Emmerich2006}.
%
% The Pareto hyper-volume has been widely used as an evaluation measure of multi-objective optimization problem. 
%
% The Pareto hyper-volume is the area of the light red region in Fig.~\ref{fig:overview}~(a) and (c), 
The Pareto hyper-volume, which is equal to the area of the light-red region in Fig.~\ref{fig:overview}~(a) and (c), is widely used as the quality measure in the multi-objective optimization.
The larger increase in the hyper-volume indicates a greater improvement in the solution.
% By using predictive distribution of GP, we can evaluate the expected increase of the Pareto hyper-volume, by which we can determine an effective candidate to sample.
%
% Therefore, by selecting the largest EIPV, we can rapidly approach to the Pareto optimal materials. 
%
An important property of the EIPV is that it can incorporate uncertainty of the prediction, due to which the balance of so-called exploit-exploration trade-off is automatically adjusted. 
%
% For example, in Fig.~\ref{fig:overview}~(c), candidate 1 has the largest expected volume because it has possibility largely improving the volume by considering its variance
% For example, in Fig.~\ref{fig:overview}~(c), candidate 1 has the largest expected volume because it has a possibility of significantly improving the volume by considering its variance, while the mean of the predictive distribution (A,B) does not largely increase the volume.
For example, in Fig.~\ref{fig:overview}~(c), although it is seen that the mean of the predictive distribution (A,B) does not largely increase the volume, candidate 1 has the largest expected volume because it has the possibility of improving the volume largely by considering its variance.

We evaluate the performance of our framework based on two case studies, wherein dynamic conductivity and static stability are simultaneously optimized.
The first case is the oxide-ion diffusivity in Er-stabilized $\delta$-Bi$_{2}$O$_{3}$ for SOFC
% \blue{ oxide-ion conductor $\rightarrow$ SOFC} 
estimated by FPMD simulations based on DFT, where solute elements and composition ratio are optimized.
The second case is the lithium-ion diffusivity in perovskite-type Li$_{3x}$La$_{2/3-x}$TiO$_{3}$ for all-solid-state LIB estimated by CMD simulations.
FPMD provides an accurate evaluation, but it is hard to collect multiple samples because of its computational difficulty.
In contrast, the CMD simulation renders it relatively easier to collect samples, even though its accuracy depends on the empirical parameters. 
For both cases, we show that our framework can explore candidate materials significantly faster than the na{\"i}ve random search.

\clearpage

% --------------------------------------------------
\section{Method} 
\label{sec:method}

% --------------------------------------------------
\subsection{Computational Methods for Physical Properties}
\label{ssec:computational-methods}

We first describe our computational models of first-principles molecular dynamics (FPMD) and classical molecular dynamics (CMD) simulations for the stability and conductivity evaluations.

% --------------------------------------------------
\subsubsection{Total energy of stable configurations}

All DFT calculations were performed using the plane-wave basis projector-augmented wave (PAW) method implemented in the VASP code \cite{Blochl1994,Kresse1999,Kresse1993,Kresse1996}.
%(1) Bl\"ochl, P. E. Projector Augmented-Wave Method. Phys. Rev. B: Condens. Matter Mater. Phys. 1994, 50, 17953$B!](B17979. 
%(2) Kresse, G.; Joubert, D. From Ultrasoft Pseudopotentials to the Projector Augmented-Wave Method. Phys. Rev. B: Condens. Matter Mater. Phys. 1999, 59, 1758. 
%(3) Kresse, G.; Hafner, J. Ab Initio Molecular Dynamics for Liquid Metals. Phys. Rev. B: Condens. Matter Mater. Phys. 1993, 47, 558. 
%(4) Kresse, G.; Furthm\CID{219}ller, J. Efficiency of Ab-Initio Total Energy Calculations for Metals and Semiconductors Using a Plane-Wave Basis Set. Comput. Mater. Sci. 1996, 6, 15$B!](B50. 
%
The exchange-correlation term was treated with the Perdew-Burke-Ernzerhof functional \cite{Perdew1996}.
%(5)
%(5) Perdew, J. P.; Burke, K.; Ernzerhof, M. Generalized Gradient Approximation Made Simple. Phys. Rev. Lett. 1996, 77, 3865$B!](B3868. 
%
The plane-wave cutoff energy was 400 eV for structure optimization and 300 eV for the FPMD simulations for computational economy.
% The plane-wave cutoff energy was 400 eV, except for the FPMD simulations, which were made with a cutoff energy of 300 eV for computational economy. 
%
% \blue{($B:o=|(B)
% The 6s and 6p electrons for Bi and the 2s and 2p electrons for O were treated as valence electrons.
% The $6s$ and $6p$ electrons for Bi and the $2s$ and $2p$ electrons for O were treated as valence electrons. 
% }
%
The energies and diffusion coefficients were estimated using 2$\times$2$\times$2 supercells of fluorite unit cell as reported the previous study \cite{Shitara2017}.

CMD simulations were performed using the pair potentials with effective partial charges.
Short-range interactions of the potentials have the Buckingham form, and long-range electrostatic interactions were calculated using the Ewald sum.
The total potential energy is given by
\begin{equation}
 V(r_{ij})=A \exp \left( - \frac{r_{ij}}{\rho}  \right) -\frac{C}{r_{ij}^{6}} + \frac{z_{i}z_{j}e^{2}}{4\pi \epsilon_{0}r_{ij}},
\end{equation}
where $z_{i}$ is the effective partial charge,  $r_{ij}$ is the interionic distance,  and $A$, $\rho$, and $C$ are parameters specific to the pairs of interacting species.
The values of these parameters were taken from the previous publication \cite{Chen1998}.

% --------------------------------------------------
\subsubsection{Diffusion coefficient}

The oxide-ion diffusivity was estimated by 
FPMD
% first-principles molecular dynamics (FPMD) 
simulations based on DFT using 80-atom cells, and the lithium-ion diffusivity was estimated by classical molecular dynamics (CMD) simulations using 1,056-atom cells. 
All MD simulations were performed with the NVT ensemble.
%% \item $B29EY0lDj(B. ($BMW3NG'(B) The temperature was controlled by the scaling of particle velocities.
%
High simulation temperature of 1600 K for FPMD and 1000 K for CMD were selected to save computational time. 
The integration time was 2 fs for both FPMD and CMD.
Mean square displacements (MSD) were calculated from NVT trajectories as, ${\rm MSD}=\langle | r_{i}(0)-r_{i}(t)|^{2} \rangle$ where $r_{i}$ is the position of particle $i$, and $r_{i}(0)$ and $r_{i}(t)$ are the positions of the particle at time 0 and time $t$, respectively. 
The diffusion coefficient $D$ in cm$^2$/s can be calculated from MSD according to the Einstein diffusion equation: $D=\frac{1}{6} \lim _{t \rightarrow \infty}\frac{d}{dt} \langle | r_{i}(0)-r_{i}(t)|^{2} \rangle$.
In FPMD, the total simulation time was 50 ps, and the first 4 ps were removed from the analysis of the diffusivity as they were the thermal equilibration steps. 
%
% In FPMD, the total simulation time was 50 ps, and the first 4 ps were removed from the analysis of the diffusivity as the thermal equilibration steps. 
% 2.5 
% 100
In CMD, the total simulation time was 100 ps, and the first 5 ps were removed because of the same reason as FPMD.

% --------------------------------------------------
\subsection{Searching Pareto Optimal using Gaussian Process}
\label{ssec:EIPV}

Suppose that there exist $n$ candidate materials, and the $i$-th material is represented by a $d$-dimensional descriptor vector $\*x_i \in \RR^d$.
Let $D_i$ and $E_i$ be 
the 
% mean squared displacement (MSD) 
diffusion coefficient and energy of the $i$-th material, respectively.
Defining
$\*y_i^\top = (y_{i,1}, y_{i,2}) = (D_i, - E_i)$,
we consider our problem as a multi-objective maximization problems with two objectives.
When 
% $y_{i,1} \leq y_{j,1}$ 
$y_{i,1} \geq y_{j,1}$ 
and 
% $y_{i,2} \leq y_{j,2}$
$y_{i,2} \geq y_{j,2}$
hold simultaneously, we write 
% $\*y_i \preceq \*y_j$ 
$\*y_i \succeq \*y_j$ 
and say ``$\*y_i$ dominates $\*y_j$''.
\emph{Pareto set} is a set of $\*y_i$ that is not dominated by any other $j \neq i$:
\begin{align*}
 P(\cY) = 
 \cbr{ \*y_i \in \cY \mid \*y_j 
 % \npreceq 
 \nsucceq
 \*y_i, 
 \forall j \neq i},
\end{align*}
where $\cY = \cbr{ \*y_1, \ldots, \*y_n }$.
Each one of $\*y \in P(\cY)$ is called the \emph{Pareto optimal} point.
For $\*y \in \cY$, if $\*y \notin P(\cY)$, at least one of 
% MSD 
the diffusion coefficient or energy can be improved without sacrificing the other property.
The boundary between the regions that are dominated and are not dominated by the current observations
% the dominated region and the region which is not dominated by the current observations 
is called the Pareto frontier, as illustrated by the red-dashed line in Fig.~\ref{fig:overview}~(a) and (c).

Let $\tilde{\cY} \subseteq \cY$ be a set of $\*y_i$ that are already computed.
To evaluate quality of $\tilde{\cY}$, we employ \emph{Pareto hyper-volume} \cite{Emmerich2006}, which is equal to the area of the light red region illustrated in Fig.~\ref{fig:overview}~(a).
% Figure~\ref{fig:hypervolume}.
%
According to a review by Riquelme et al. \cite{Riquelme2015}, the Pareto hyper-volume is the most widely used optimality criterion in the multi-objective optimization. 
To define the volume, a reference point $\*y_{\mathrm{ref}}$ (black square in Figure~\ref{fig:overview}~(a)) is required, which is a point dominated by all the other points ($\*y_{\mathrm{ref}} \preceq \*y$ for $\forall \*y \in \cY$).
%  % $\*y_{\mathrm{ref}} \succeq \*y$ 
%
In practice, for each dimension of the reference point $\*y^{\mathrm{ref}}$, we set $y^{\mathrm{ref}}_{i} = \min_{j \in \cS} y_{i,j} - 0.1 r_i$, where $r_i$ is the value range of the $i$-th dimension of $\*y$ in the already sampled points.
The Pareto hyper-volume is mathematically defined as the area of the region in which any point satisfies the following two conditions: (1) being dominated by at least one of $\*y$ in $P(\tilde{\cY})$, and (2) dominating the reference point.
% , as illustrated in the light red region in Fig.~\ref{fig:overview}~(a).
%
% Figure~\ref{fig:hypervolume} shows an illustration of Pareto hypervolume.
% Based on the reference point, Pareto hypervolume is defined by the shaded area in Figure~\ref{}, which is determined by the area dominated by $P(\tilde{\cY})$.
% $  \{ \*y \mid \*y \preceq \*y', \exists \*y' \in P(\cY), \*y_{\mathrm{ref}} \preceq \*y \}$
% 
Let
\begin{align*}
 \mathrm{Vol}(P(\tilde{\cY})),
\end{align*}
be the hyper-volume of the Pareto set $P(\tilde{\cY})$.
Note that $\mathrm{Vol}(P(\tilde{\cY}))$ is non-decreasing with respect to addition of an element into $\tilde{\cY}$, and attains the maximum value when $\tilde{\cY}$ contains all the Pareto optimal points $P(\cY)$.
Since 
$\mathrm{Vol}(P(\tilde{\cY}))$
is maximized only when all the Pareto optimal points are included in $\tilde{\cY}$, the Pareto set identification problem can be seen as the maximization of $\mathrm{Vol}(P(\tilde{\cY}))$.

% We employ the Gaussian process (GP) model for building prediction models of MSD and energy.
%
By using the Gaussian process (GP) model, the diffusion coefficient and negative energy of the $i$-th material are represented by the two Gaussian random variables $f^{(D)}_i$ and $f^{(E)}_i$, respectively.
% MSD and negative energy are represented by the two Gaussian random variables $\*f^{\mathrm{(D)}}$ and $\*f^{\mathrm{(E)}}$ respectively, which are defined by 
%
% Given the computed $\tilde{\cY}$, the conditional distributions provide predictive distributions for candidate crystal structures:
For the given computed $\tilde{\cY}$, we write the predictive Gaussian distributions for candidate materials as
\begin{align*}
 f^{(D)}_i \mid \tilde{\cY} &\sim \cN( \mu_D(\*x_i), \sigma_D^2(\*x_i) ), \\ %^2 
 f^{(E)}_i \mid \tilde{\cY} &\sim \cN( \mu_E(\*x_i), \sigma_E^2(\*x_i) ),
\end{align*}
where
% $\mu_D, \mu_E: \RR^d \rightarrow \RR$
$\mu_D(\*x)$ and $\mu_E(\*x)$
are conditional mean functions, and
% $\sigma^2_D, \sigma^2_E: \RR^d \rightarrow \RR$
$\sigma^2_D(\*x)$ and $\sigma^2_E(\*x)$
are conditional variance functions.
%
% See e.g., \cite{Rasmussen2006} for further detail of GP.
Further details of GP are given in \cite{Rasmussen2006}.
These predictive distributions are illustrated as the green vertical distributions in Fig.~\ref{fig:overview}~(b), and the green circles in Fig.~\ref{fig:overview}~(c).

From the GP models, we estimate the possible increase of the Pareto hyper-volume when new observations of 
% MSD 
diffusion coefficient
and energy are obtained.
Considering the expectation of GP predictions, we obtain the expected increase of Pareto hyper-volume (EIPV):
% To select a next point to be sampled, we consider the expected improvement in Pareto hypervolume (EIPV):
%
\begin{align*}
 \mathrm{EIPV}(\*x_i) =
 % \EE_{p(\*y_i \mid \tilde{\cY})} 
 \EE_{p(\*f_i \mid \tilde{\cY})} 
 \left[
 % \max\left(
 % \mathrm{Vol}(P(\cY \cup \*y_i))
 \mathrm{Vol}(P(\cY \cup \*f_i))
 -
 \mathrm{Vol}(P(\cY))
 % 0 \right)
 \right],
\end{align*}
where $\*f_i = [f_i^{(D)}, f_i^{(E)}]^\top$.
% where each dimension of $\*y_i \mid \tilde{\cY}$
% follows
% $\cN(\mu_D(\*x_i), \sigma_D^2(\*x_i) + \varepsilon)$ 
% and
% $\cN(\mu_E(\*x_i), \sigma_E^2(\*x_i) + \varepsilon)$
% with the noise variance parameter $\varepsilon$.
%
The calculation of EIPV can be performed analytically \cite{Emmerich2006}.
Since diffusion coefficient and energy can have different scales, we calculate volumes after standardizing the observed $\*y$ so that each dimension has $0$ mean and unit variance.
We iteratively select $\argmax_i \mathrm{EIPV}(\*x_i)$ as the next candidate.
EIPV is a generalization of the well-known expected improvement in the single objective optimization of Bayesian optimization \cite{Shahriari2016}.
%
% By considering the expectation over the possible increase of the volume, EIPV measures how much volume improvement can be achieved 
% 
% By considering the expectation over the possible increase of the volume, uncertainty of the prediction can be taken into consideration.
In view of the expectation over the possible increase of the volume, uncertainty of the prediction can be taken into consideration.
For example, in candidate 1 of Fig.~\ref{fig:overview}~(c), although the predictive mean is close to the Pareto frontier, this candidate has the possibility of large increase of volume when the variance of the prediction is considered.

\clearpage

% --------------------------------------------------
\section{Results}
\label{sec:result}

% --------------------------------------------------
% \clearpage
% \subsection{Case Studies}

\begin{figure}[t]
 \centering
 \subfloat[$\delta$-Bi$_2$O$_3$]{
 \includegraphics[clip,width=0.2\tw]{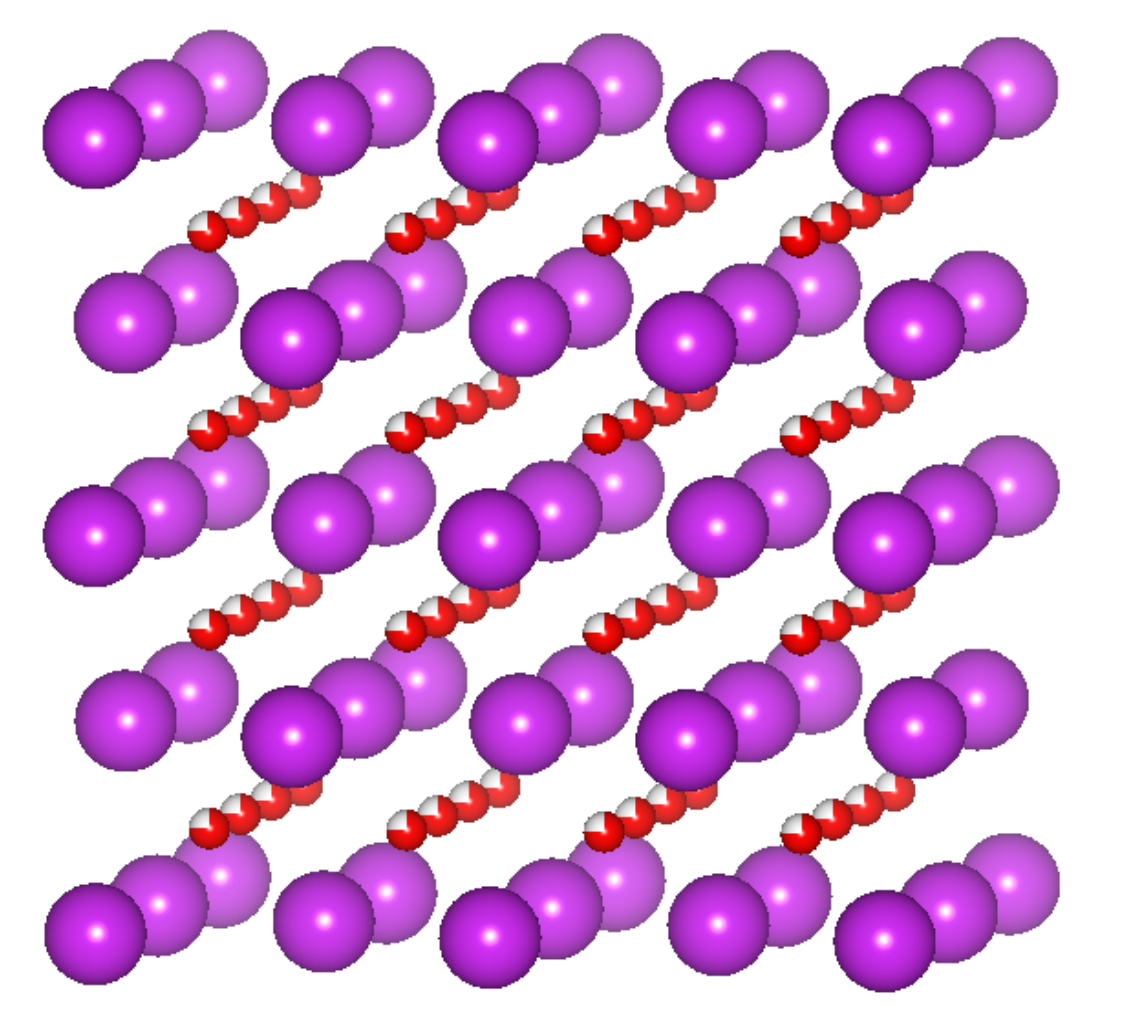}   
 }
 \subfloat[Li$_{3x}$La$_{2/3-x}$TiO$_{3}$]{
 \includegraphics[clip,width=0.2\tw]{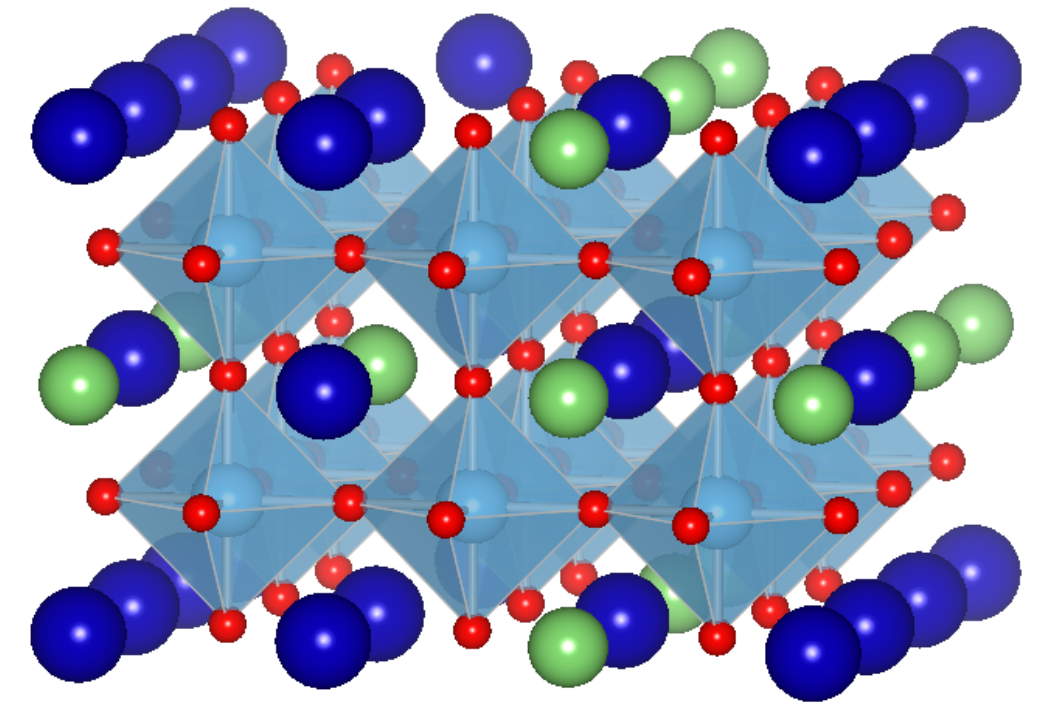}   
 }
 \caption{
 Atomic configurations of (a) $\delta$-Bi$_{2}$O$_{3}$ and (b) Li$_{3x}$La$_{2/3-x}$TiO$_{3}$. 
 A red ball denotes an O atom, a pink one a Bi atom, a light blue one a Ti atom, a blue one a La atom, and a green one a Li atom. 
 The number of O atoms in $\delta$-Bi$_{2}$O$_{3}$ is adjusted in order to maintain charge neutrality. 
 }
 \label{fig:structures}
\end{figure}

We demonstrate efficiency of our framework through case studies on 
\ce{Bi2O3} and La$_{2/3-x}$Li$_{3x}$TiO$_{3}$, which are illustrated in Fig.~\ref{fig:structures}.
For the kernel function in the GP models, which determines similarity between two materials, 
% In the both GP models for conductivity and stability, 
we used the standard Gaussian kernel $k(\*x_i,\*x_j) = \exp(- \| \*x_i - \*x_j \|^2 / \gamma)$, where $\gamma$ is fixed by the average squared distance value between candidate materials.
For both scenarios, we first generate $2$ initial points randomly.
Here, we compared the performance of EIPV with a simple random sampling.

% --------------------------------------------------
\subsection{Case Study I: Oxygen Diffusion in Bi$_{2}$O$_{3}$}

% --------------------------------------------------
% \subsection{Problem Setting}

The supercell included 32 cations and the cation arrangement was assumed to be disordered.
We considered the composition of Bi$_{1-x-y-z}$Er$_{x}$Nb$_{y}$W$_{z}$O$_{48+y+3/2z}$, where the number is 16 atoms or more for Bi, and 64 atoms or less for O. 
The number of the compositions is 335. 
Disordered arrangements were made as special quasi-random structures (SQS) \cite{Zunger1990} of the anion and cation sublattices considering pairwise interactions of all of the anion-cation, the anion-anion, and cation-cation pairs.
% Disordered arrangements were made as special quasi-random structures (SQS)(7) of the anion and cation sublattices considering pairwise interactions of all of the anion$B!](Bcation, the anion$B!](Banion, and cation-cation. 
%(7) Zunger, A.; Wei, S.; Ferreira, L.; Bernard, J. Special Quasirandom Structures. Phys. Rev. Lett. 1990, 65, 353$B!](B356. 
%
They were constructed by simulated annealing implemented in CLUPAN code \cite{Seko2009,Seko2010}. 
% They were constructed by simulated annealing implemented in CLUPAN code (8,9). 
%(8) Seko, A.; Koyama, Y.; Tanaka, I. Cluster Expansion Method for Multicomponent Systems Based on Optimal Selection of Structures for Density-Functional Theory Calculations. Phys. Rev. B: Condens. Matter Mater. Phys. 2009, 80, 165122. 
%(9) Seko, A. Exploring Structures and Phase Relationships of Ceramics from First Principles. J. Am. Ceram. Soc. 2010, 93, 1201$B!](B 1214. 
%
The formation energies were calculated from simple cation oxides.
% The formation energies were calculated from simple cation oxides. 
%
When multiple polymorph structures are known for the end-member oxide of the cations, the lowest energy structure was used as the reference. 
% When multiple polymorph structures are known for the end-member oxide of cations, the lowest energy structure was used as the reference. 
%
The configurational entropy was taken into account for the disordered structure by the point approximation.
% The configurational entropy was taken into account for the disordered structure by the point approximation. 

Oxygen-ion conductors have been used as oxygen sensors, solid oxide fuel cells and oxygen separation membranes \cite{Knauth2002,Azad1994}.
% (11) Knauth, P.; Tuller, H. L. Solid-State Ionics: Roots, Status, and Future Prospects. J. Am. Ceram. Soc. 2002, 85, 1654$B!](B1680.
% (12) Azad, A. M.; Larose, S.; Akbar, S. A. Bismuth Oxide-Based Solid Electrolytes for Fuel Cells. J. Mater. Sci. 1994, 29, 4135$B!](B4151.
%
Bismuth sesquioxide with a cubic form, $\delta$-Bi$_2$O$_3$, is known to have a high oxide-ion conductivity exceeding 1 S/cm at 1000 K, which is much higher than that of yttria-stabilized zirconia (YSZ), a standard oxide-ion conducting electrolyte \cite{Sammes1999,Takahashi1978}. 
% (13) Sammes,N.;Tompsett,G.;Na\CID{71}e,H.;Aldinger,F.BismuthBased Oxide Electrolytes\UTF{E0D5}Structure and Ionic Conductivity. J. Eur. Ceram. Soc. 1999, 19, 1801$B!](B1826.
% (14) Takahashi, T.; Iwahara, H. Oxide Ion Conductors Based on Bismuthsesquioxide. Mater. Res. Bull. 1978, 13, 1447$B!](B1453.
%
% We reported 
The selection of solute elements in $\delta$-Bi$_2$O$_3$ solid solutions from the systematic first-principles calculations of formation energies and diffusion coefficients are previously reported \cite{Shitara2017}.
However, the automatic exploration of solute concentrations was not discussed in \cite{Shitara2017}, and it is important challenge in practice that needs to be tackled to counter its large computational cost.
%
% In this case, concentrations of three solutes (Nb, W, and Er) are considered as descriptors $\*x \in \RR^3$. %  in order to optimize the solutes concentrations. % only
In this study, concentrations of solutes such as Nb, W, and Er, are considered as descriptors $\*x \in \RR^3$ in order for optimization. 
We prepared a dataset of formation energies at 773 K and diffusion coefficients at 1600 K for each composition using first-principles calculations. 
%
% The number of data is 335. 

\begin{figure}[t]
 \begin{center}

  \subfloat[EIPV]{
  \includegraphics[clip,width=0.24\tw]{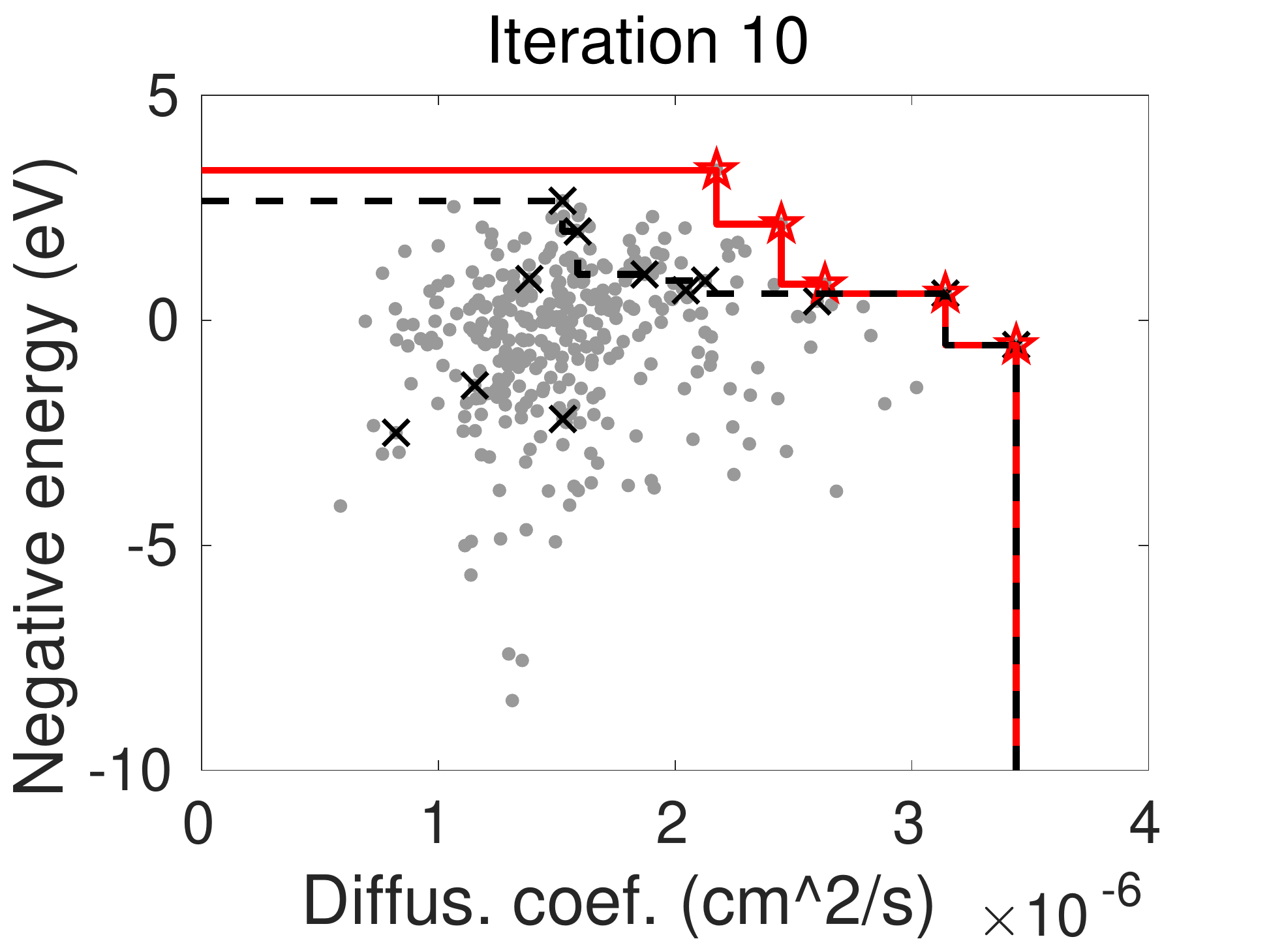} 
  \includegraphics[clip,width=0.24\tw]{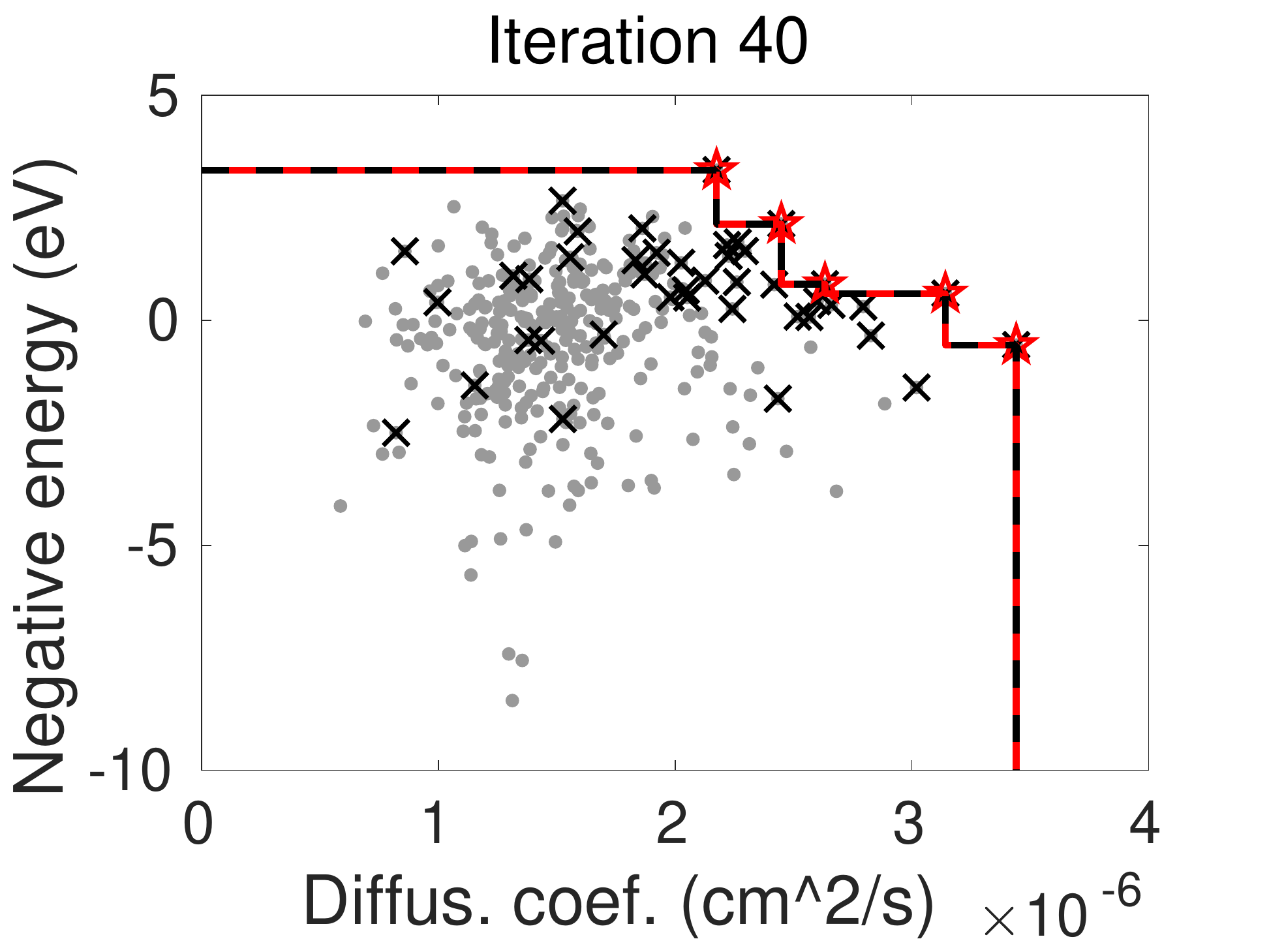} 
  \includegraphics[clip,width=0.24\tw]{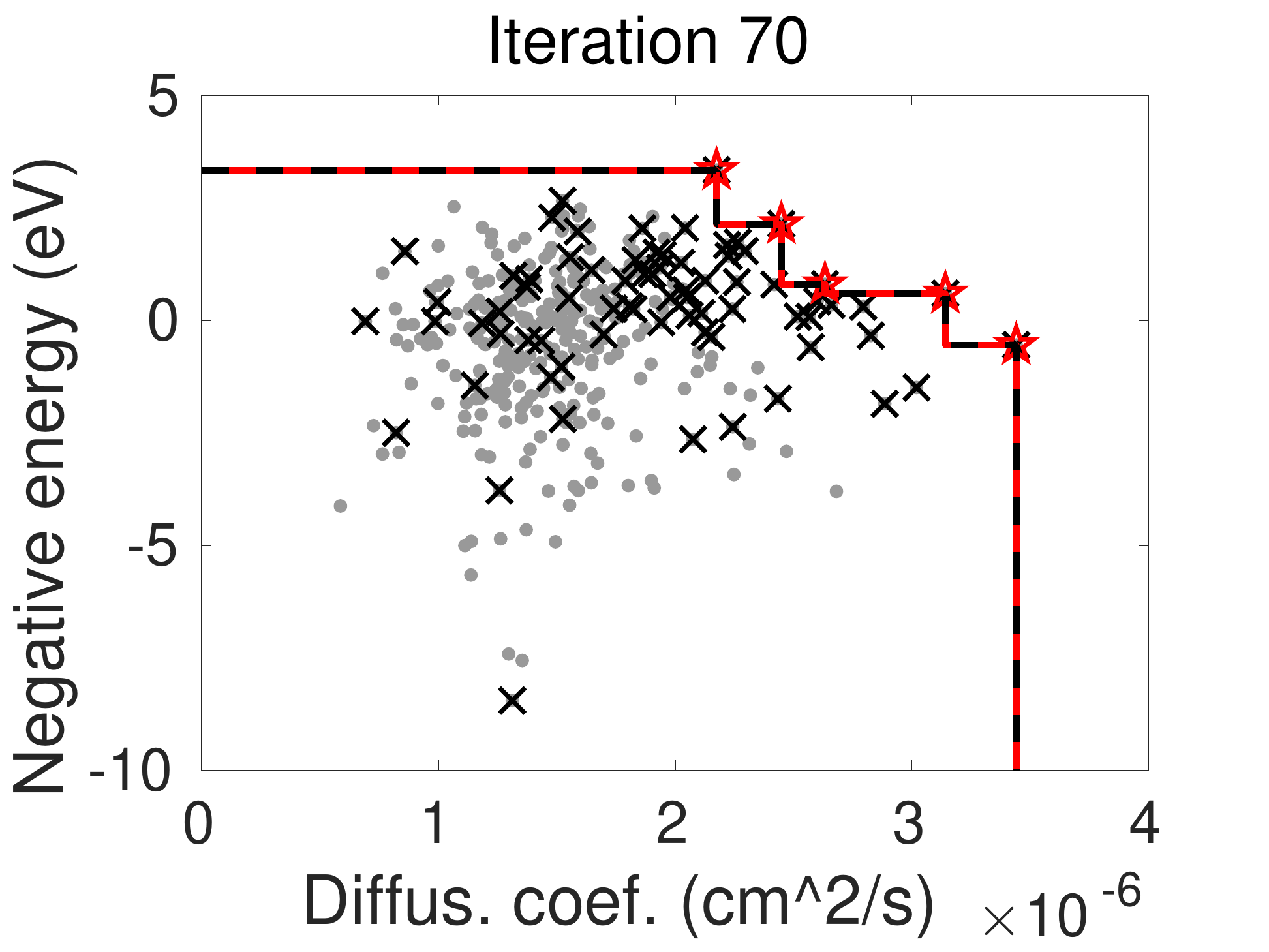} 
  \includegraphics[clip,width=0.24\tw]{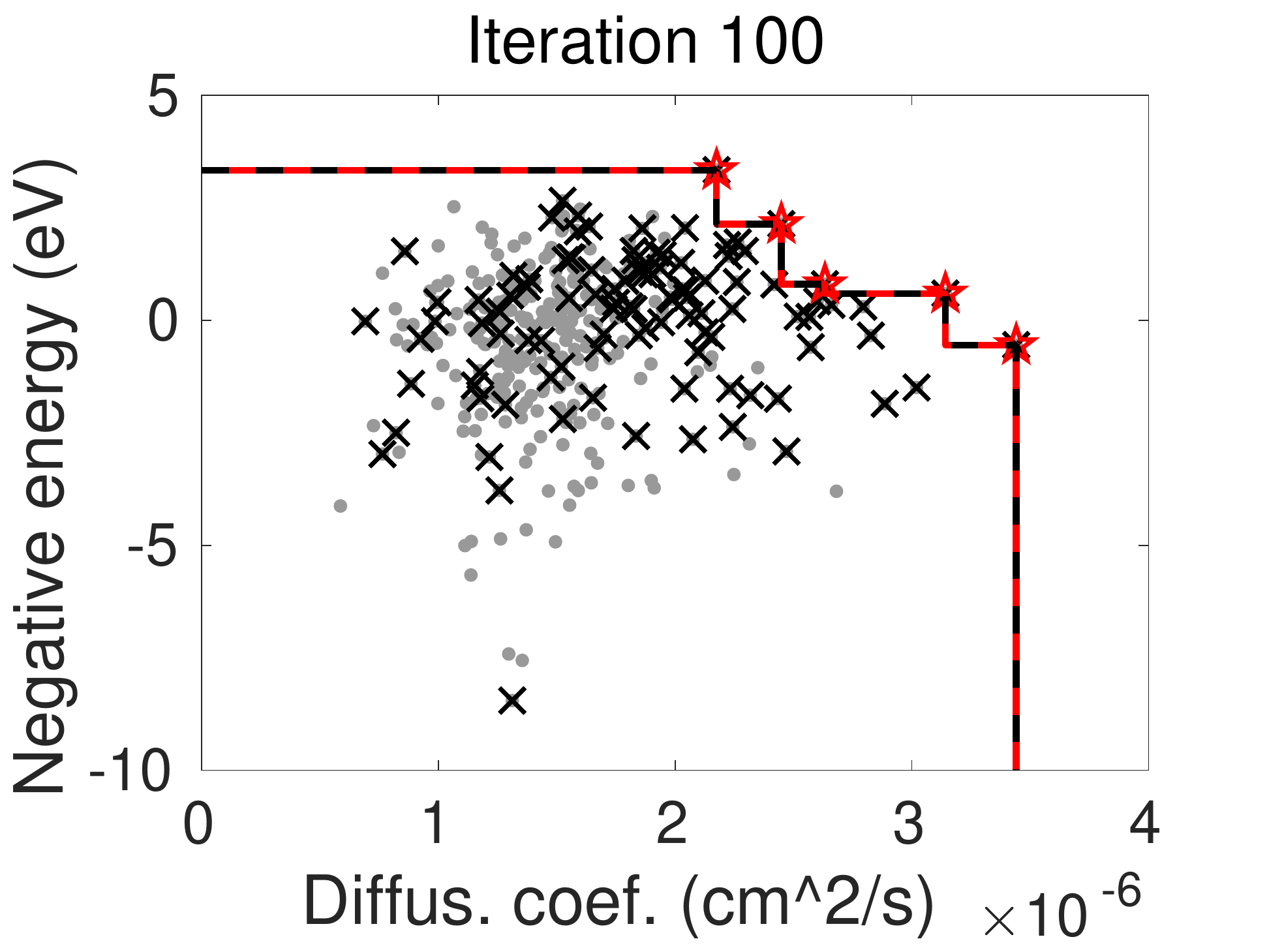} 
  }

  \subfloat[Random]{
  \includegraphics[clip,width=0.245\tw]{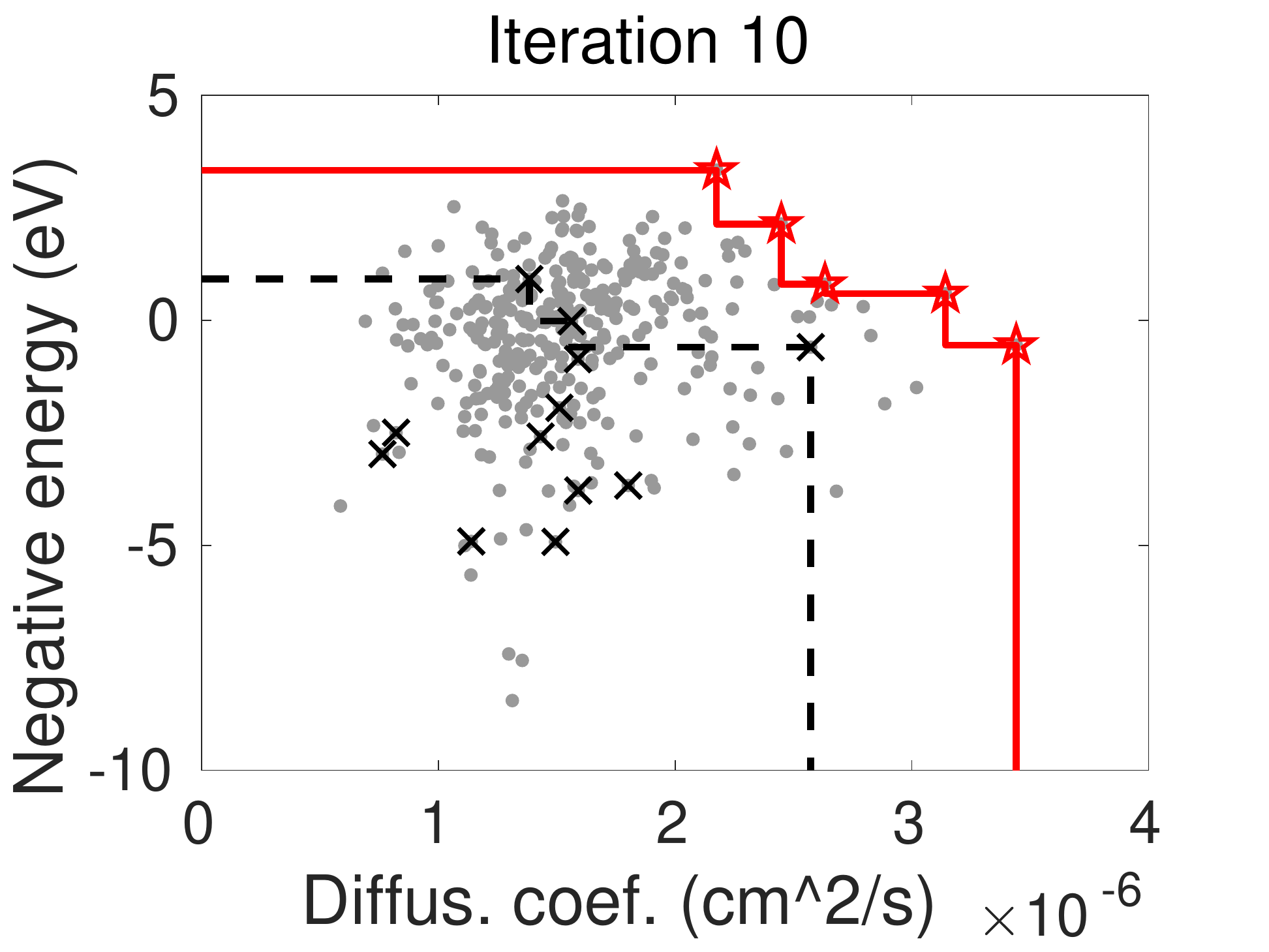}
  \includegraphics[clip,width=0.245\tw]{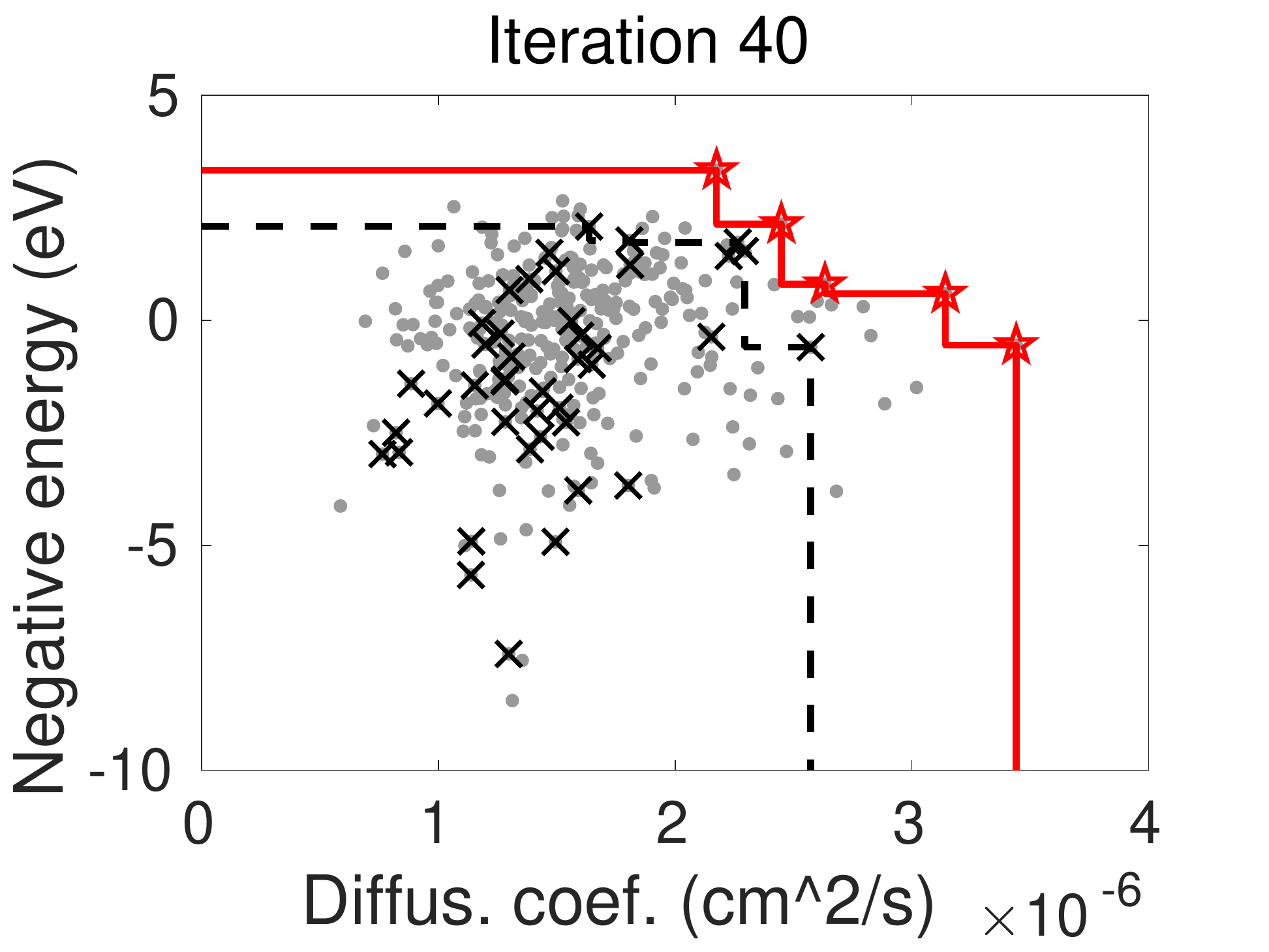}
  \includegraphics[clip,width=0.245\tw]{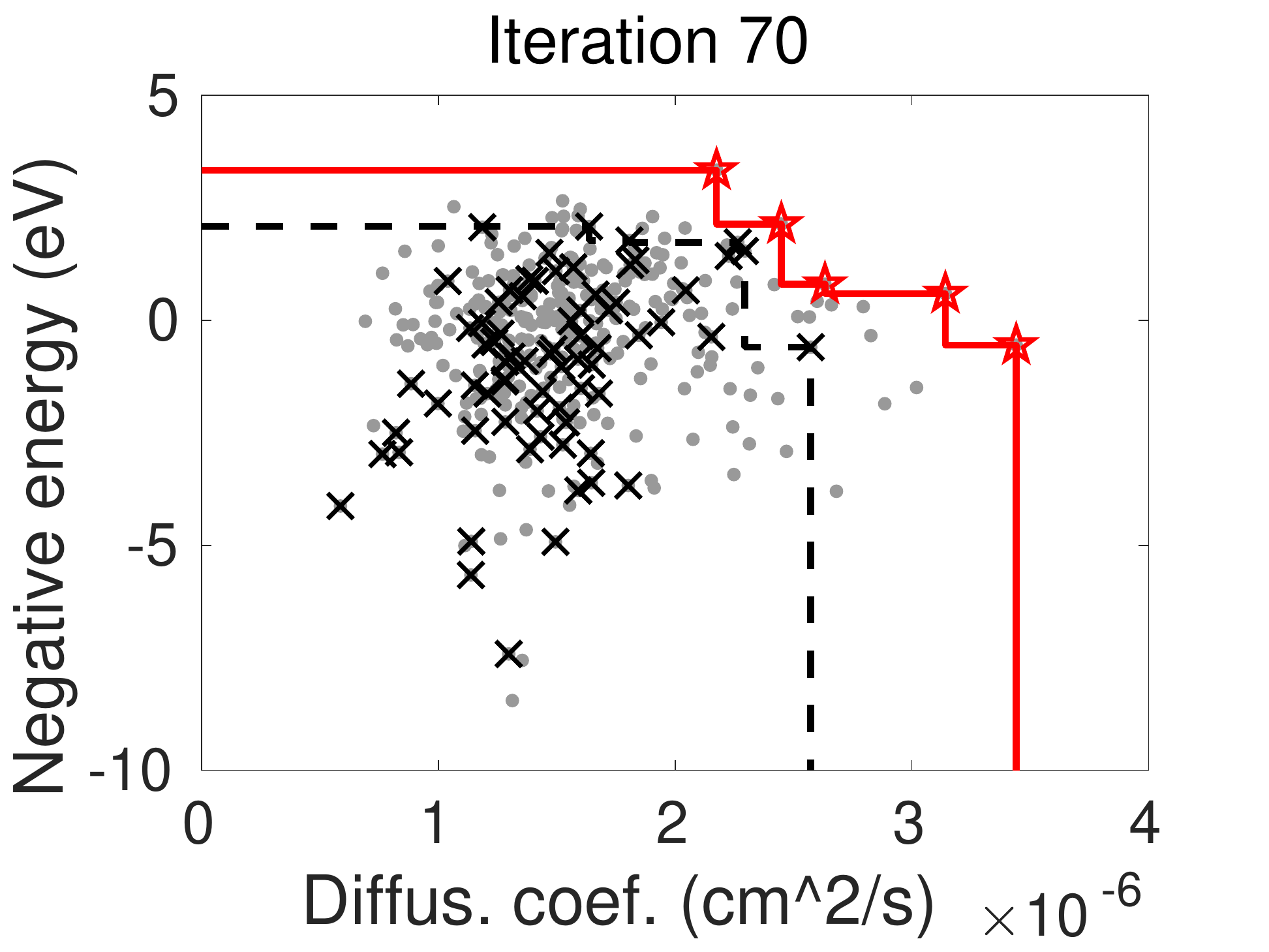}
  \includegraphics[clip,width=0.245\tw]{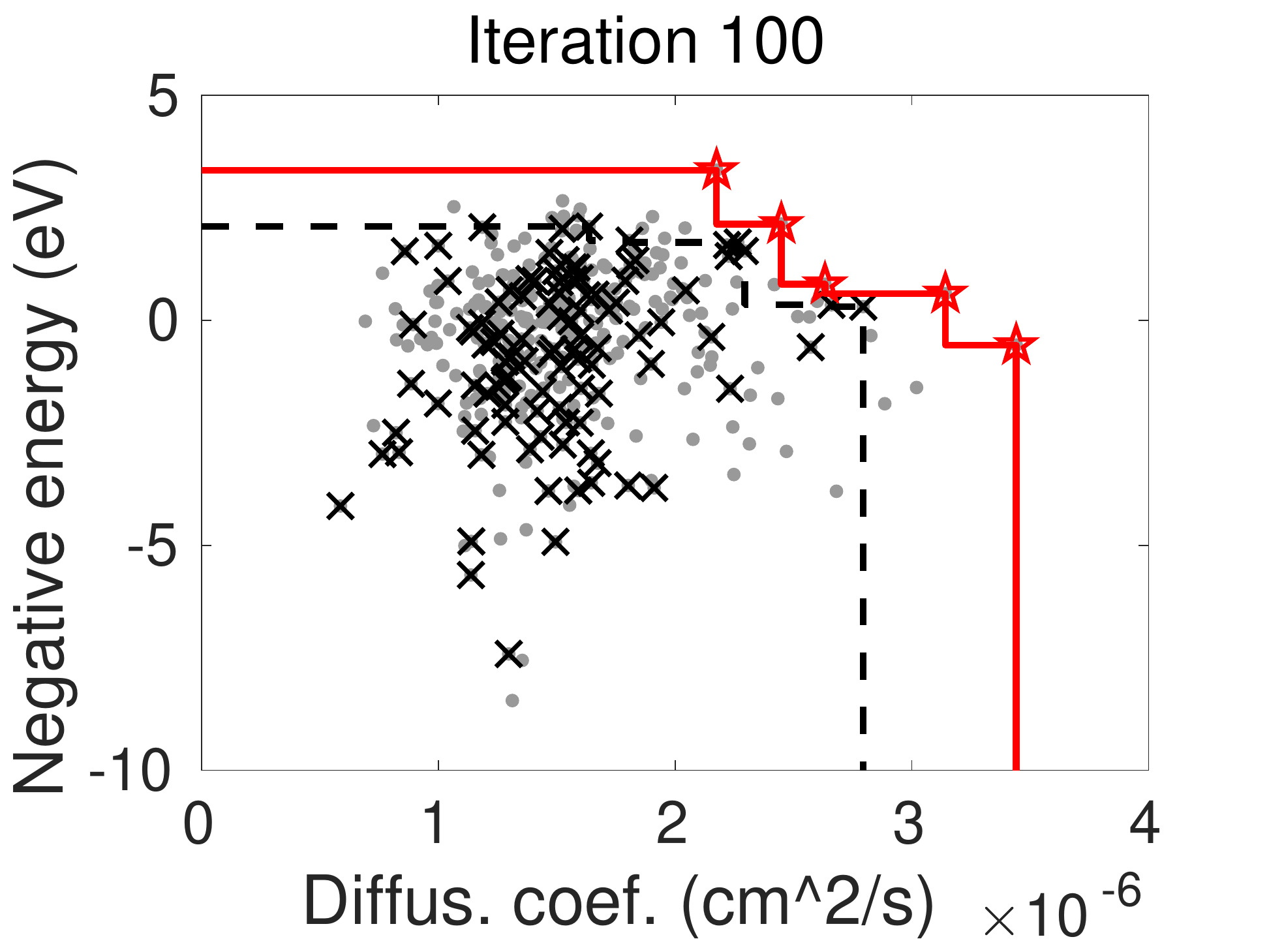} 
  }

 \end{center}
 \caption{
 An example of the optimization process for \ce{Bi2O3}.
 The vertical axis is the negative value of formation energy (eV), and the horizontal axis is the diffusion coefficient (cm$^2$/s).
 The red stars are the Pareto optimal points.
 The black crosses are sampled points for each number of iterations, and the gray dots are candidate points.
 }
 \label{fig:demo-Bi2O3}
\end{figure}

\begin{figure}[t]
 \begin{center}
  \includegraphics[clip,width=0.4\tw]{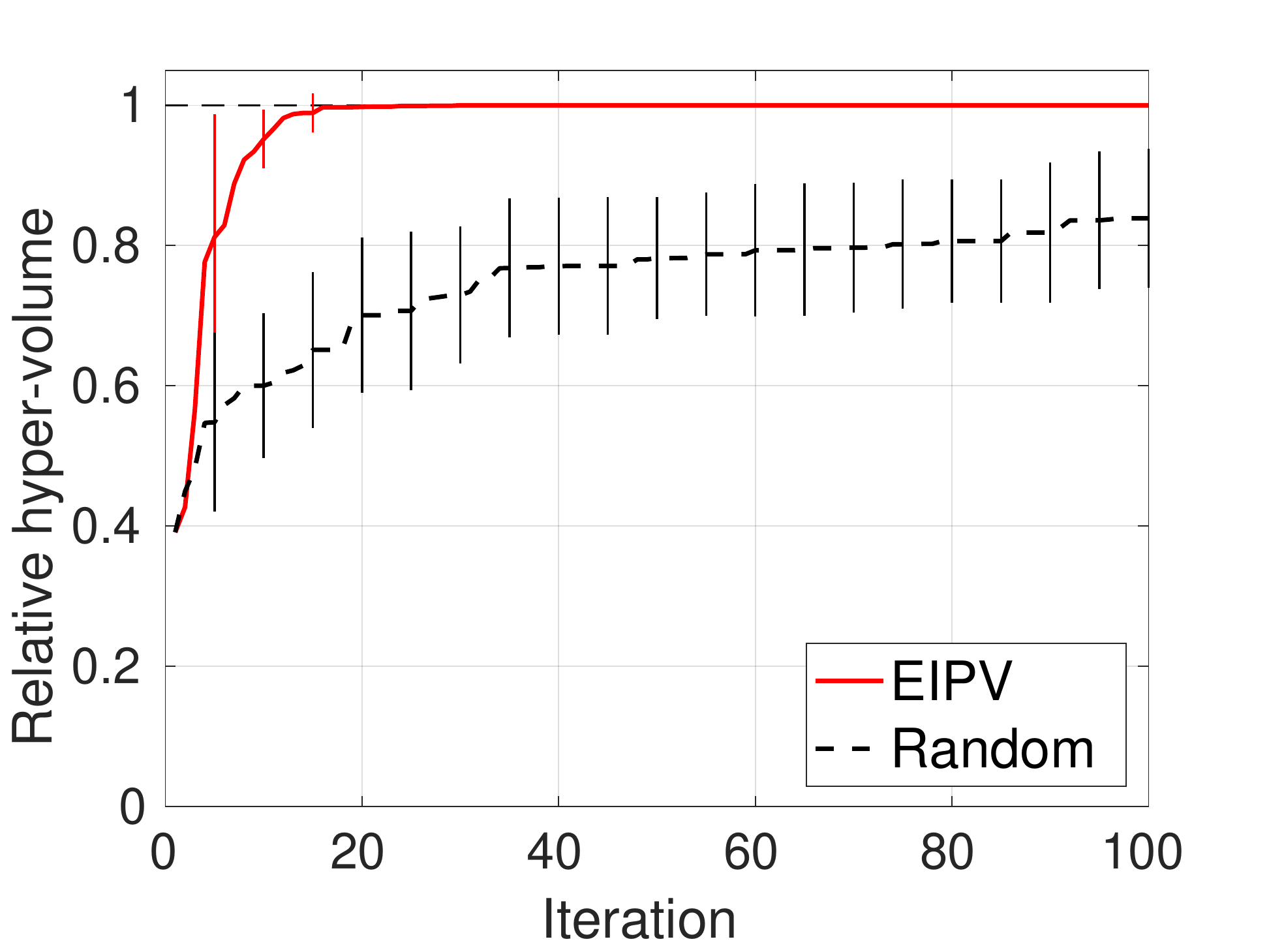} 
 \end{center}
 \caption{ 
 The transition of the relative hyper-volume for the \ce{Bi2O3} data.
 The average of $10$ runs is shown, and the vertical line is standard deviation.
 }
 \label{fig:hvr-Bi2O3}
\end{figure}

Figure~\ref{fig:demo-Bi2O3} shows the Pareto frontier at the iterations 10, 40, 70, and 100 of EIPV, and the random sampling, respectively.
We see that with only a small number of iterations, EIPV rapidly found points close to the Pareto optimal, as compared with the random sampling.
%
% \figurename~\ref{fig:demo-Bi2O3} shows (a) the initial state in which $10$ points were randomly selected, (b) the $10$-th iteration, (c) the $15$-th iteration, and (d) the $20$-th iteration.
%
% We see that with only a small number of iterations, the proposed method started to find points close to Pareto optimal.
%
Figure~\ref{fig:hvr-Bi2O3} shows the transition of the relative hyper-volume value, defined by $\mathrm{Vol}(P(\tilde{\cY})) / \mathrm{Vol}(P(\cY))$.
EIPV increased the volume substantially faster than the random search, and the standard deviation, shown as the vertical line, rapidly decreased.
We see that EIPV achieved the almost optimal value around the 20-th iteration, whereas the random sampling obtained results that were still far from the optimal even at the 100-th iteration.
% converged to the optimal around the iteration 40 robustly against the initialization difference.
%
% In contrast, the Random search was still far from the optimal even at the iteration 100.

% --------------------------------------------------
% Illustration (Bi2O3)
% --------------------------------------------------

% --------------------------------------------------
% Hyper-volume (Bi2O3)
% --------------------------------------------------

% --------------------------------------------------
\subsection{Case Study II: Li Diffusion in LLTO}

% --------------------------------------------------
% \subsection{\red{Problem Setting}}

Perovskite (Pv) type La$_{2/3-x}$Li$_{3x}$TiO$_{3}$ (LLTO) for $x=0.11$ shows the highest ion conductivity among oxide-based electrolytes \cite{Stramare2003}. 
We dealt with Li$_{3/9}$La$_{5/9} \boxempty_{1/9}$TiO$_{3}$ models, where six Li ions, ten La ions, and two vacancies are located at A sites of Pv structures, and constructed 1120 structural patterns.
For the descriptor, we simultaneously used radial distribution function (RDF) \cite{Schutt2014} and smooth overlap of atomic positions (SOAP) \cite{De2016}, both of which have been widely used to represent materials for defining the input of machine-learning models \cite{Ramprasad2017,Tamura2017}. 
%
% As a result, we produced the 2185 dimensional vector $\*x$ (84 dimensional RDF + 2101 dimensional SOAP) for each one of candidate.
We produced the 84 dimensional RDF vector $\*x^{\rm RDF}$ and the 2101 dimensional SOAP vector $\*x^{\rm SOAP}$ for each candidate.
In the kernel function calculation, we normalized these two descriptors as 
$k(\*x_i,\*x_j) = \exp(- \| \*x_i^{\rm RDF} - \*x_j^{\rm RDF} \|^2 /\gamma^{\rm RDF} - \| \*x^{\rm SOAP}_j - \*x^{\rm SOAP}_j \|^2 / \gamma^{\rm RDF} )$,
where 
$\gamma^{\rm RDF}$
and
$\gamma^{\rm SOAP}$
are the average squared distance values of RDF and SOAP in candidate materials, respectively.
Note that these calculations do not incur a large computational cost when compared with stability and conductivity evaluation.

\begin{figure}[t]
 \begin{center}

  \subfloat[EIPV]{
  \includegraphics[clip,width=0.24\tw]{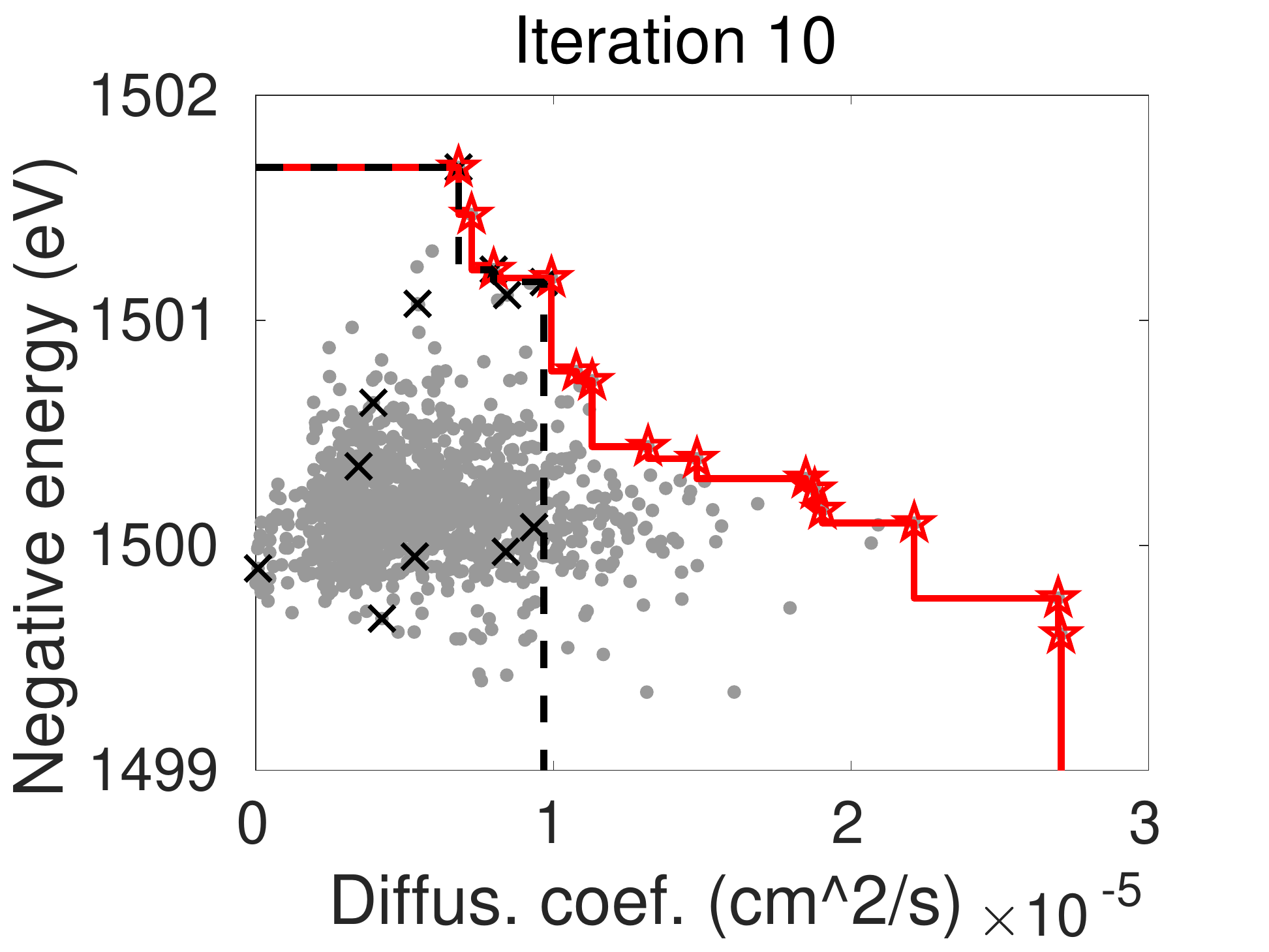} 
  \includegraphics[clip,width=0.24\tw]{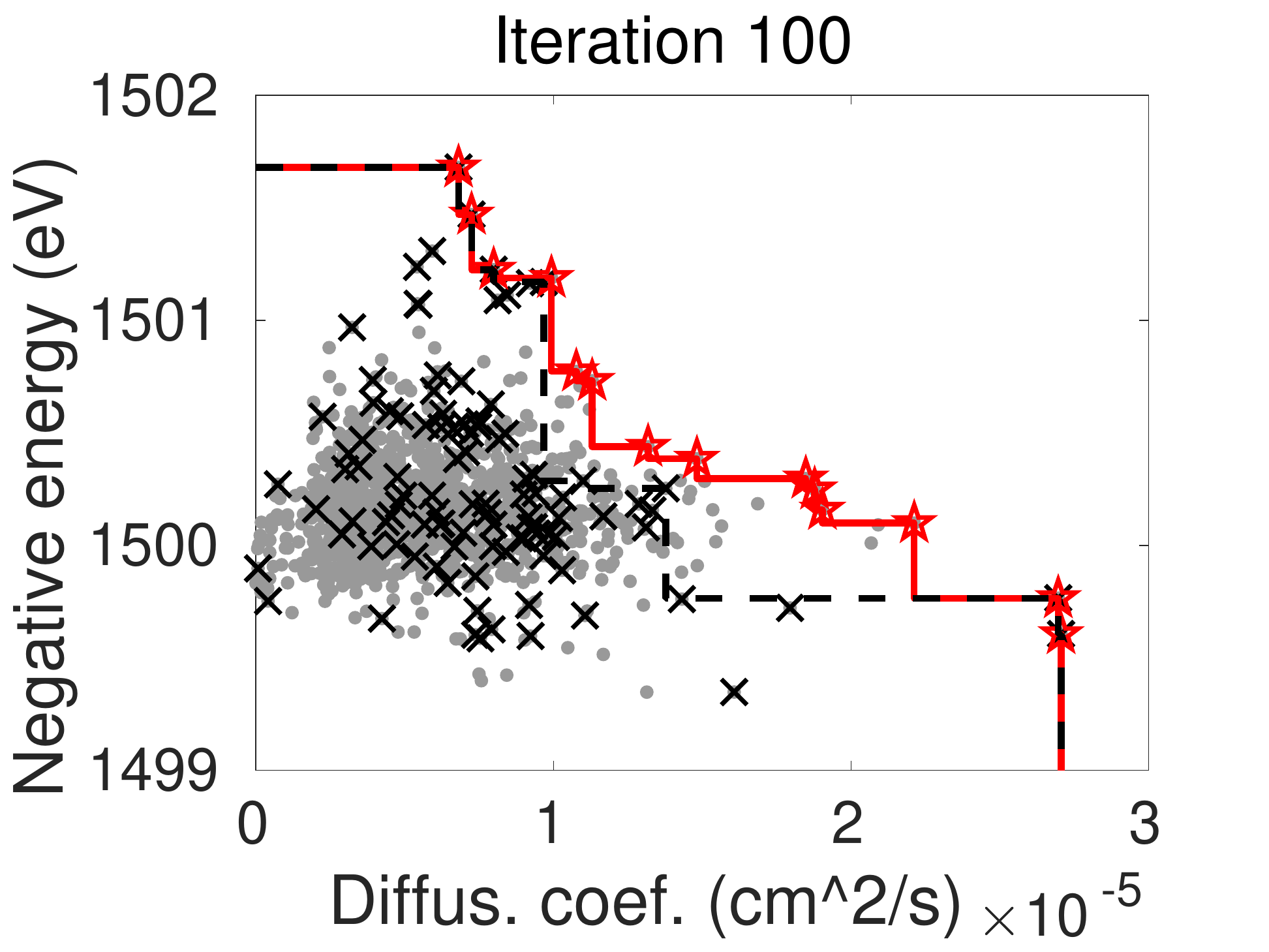} 
  \includegraphics[clip,width=0.24\tw]{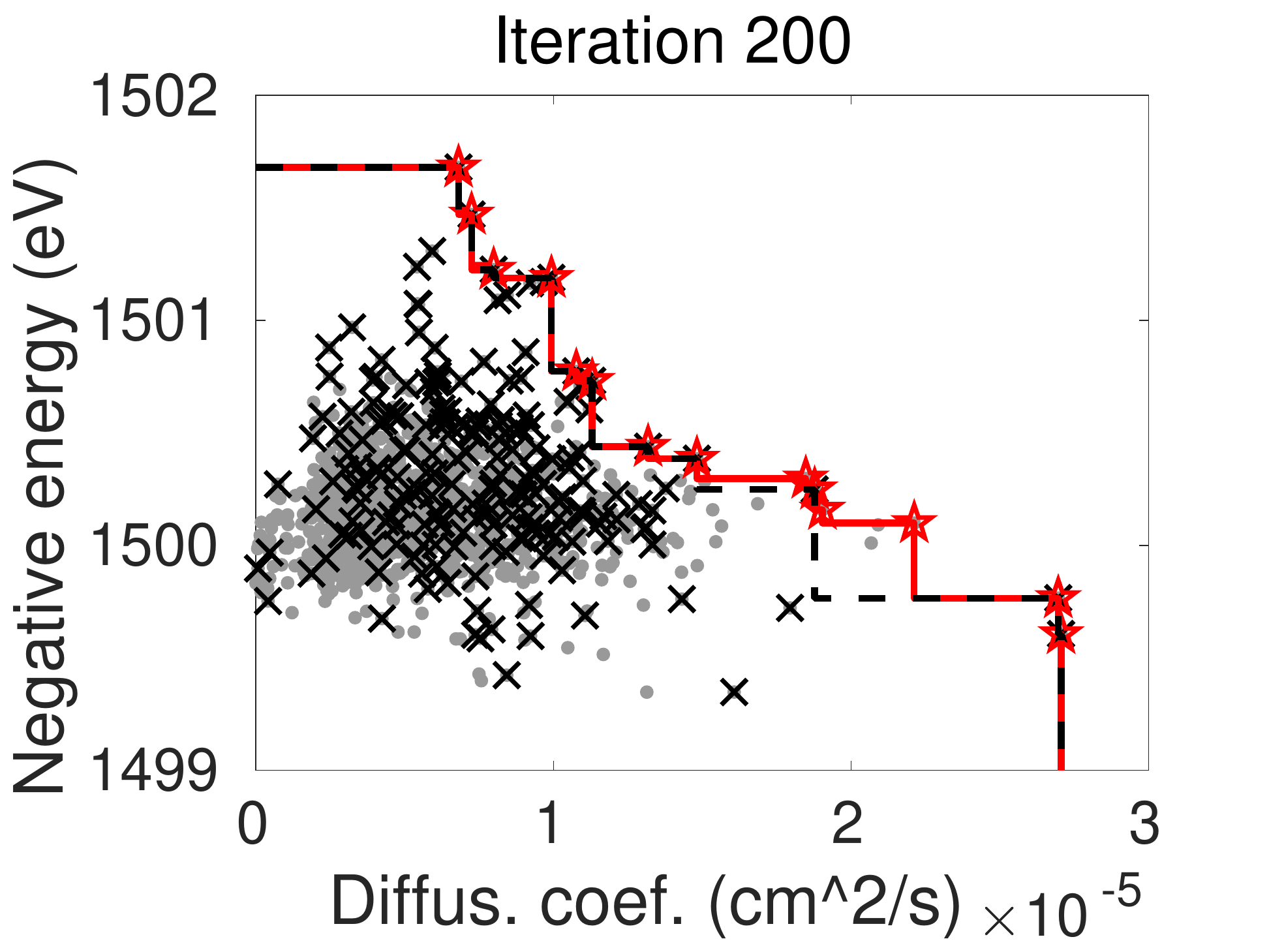} 
  \includegraphics[clip,width=0.24\tw]{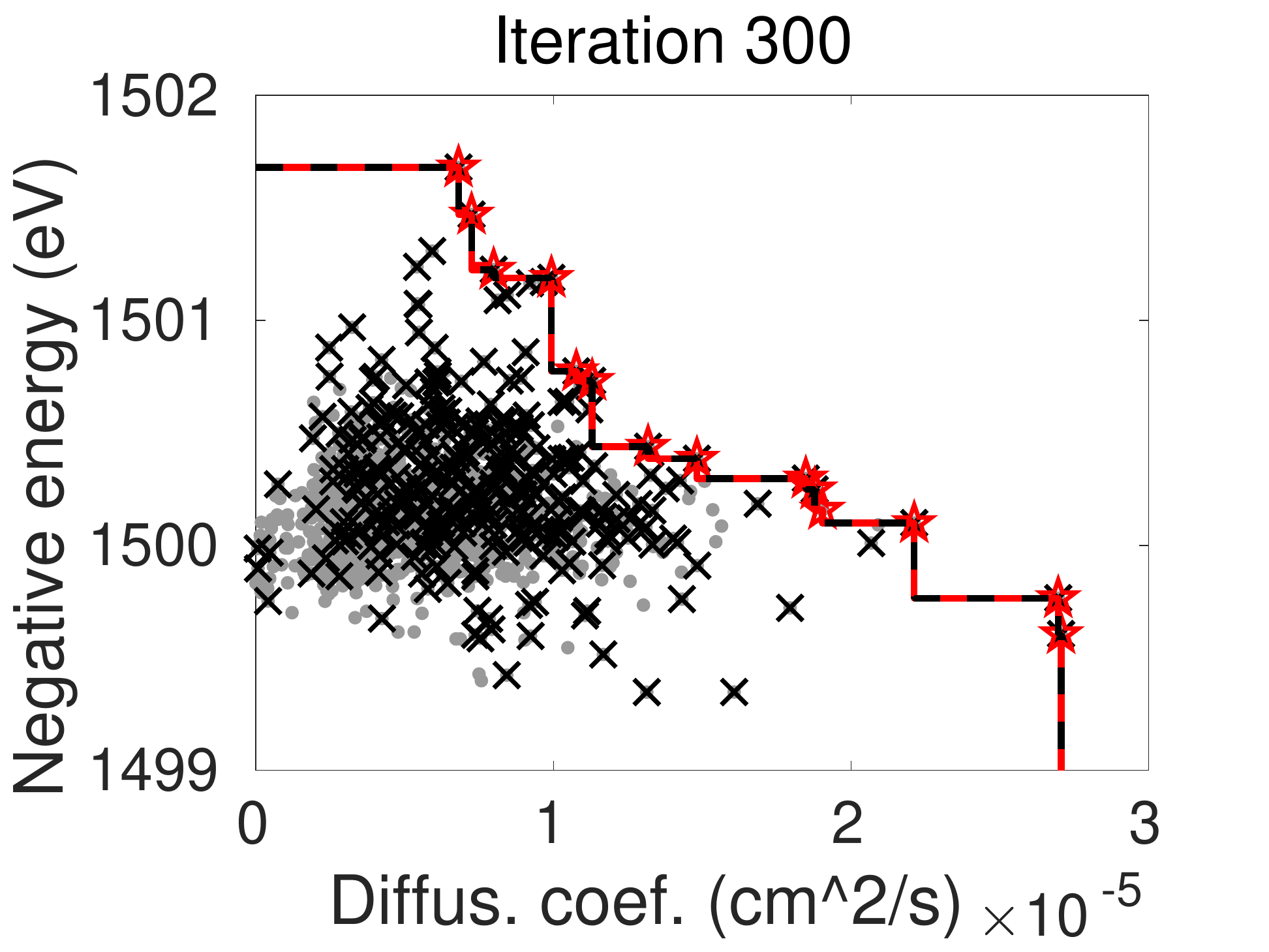} 
  }
 
  \subfloat[Random]{
  \includegraphics[clip,width=0.24\tw]{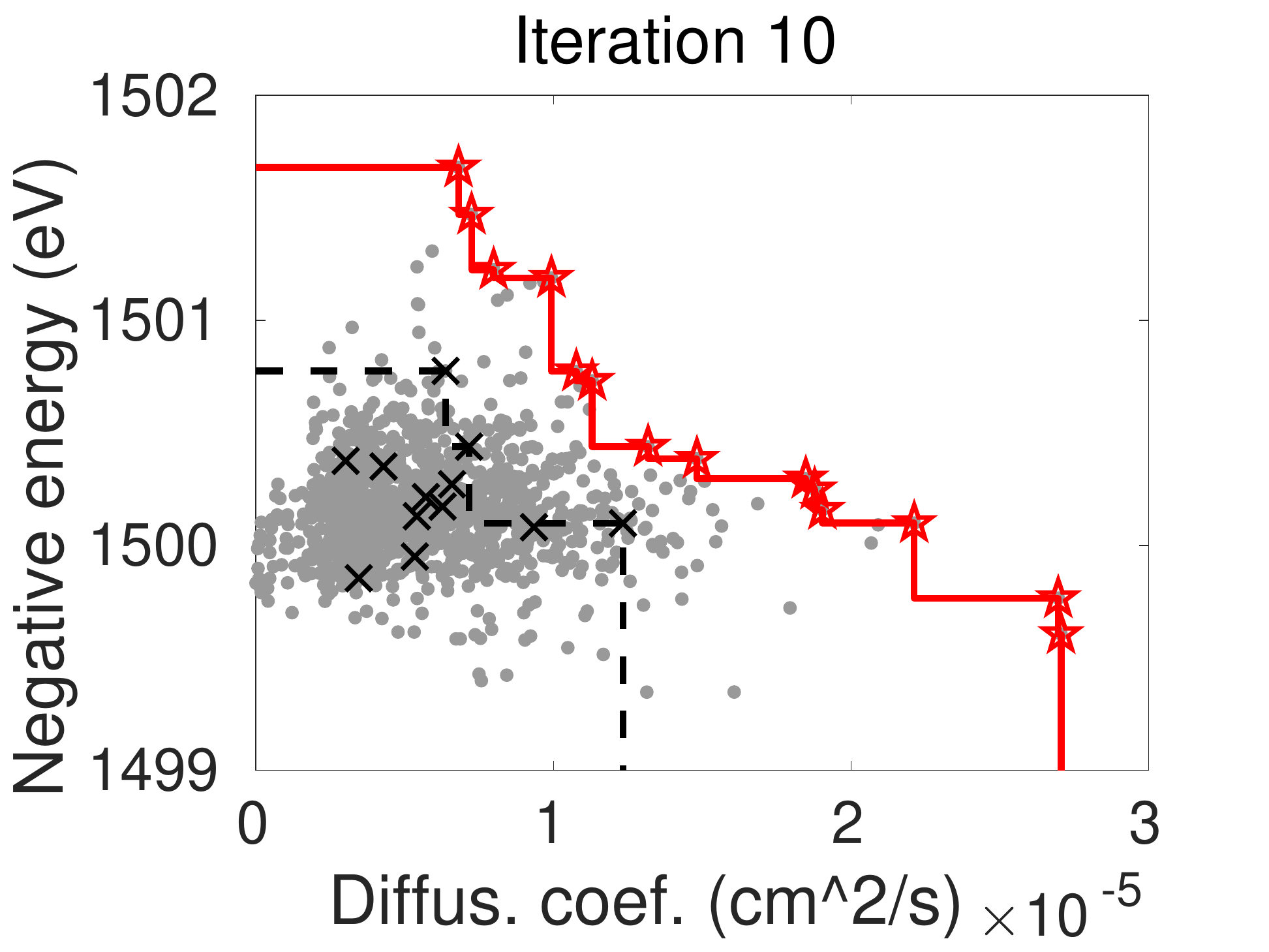}
  \includegraphics[clip,width=0.24\tw]{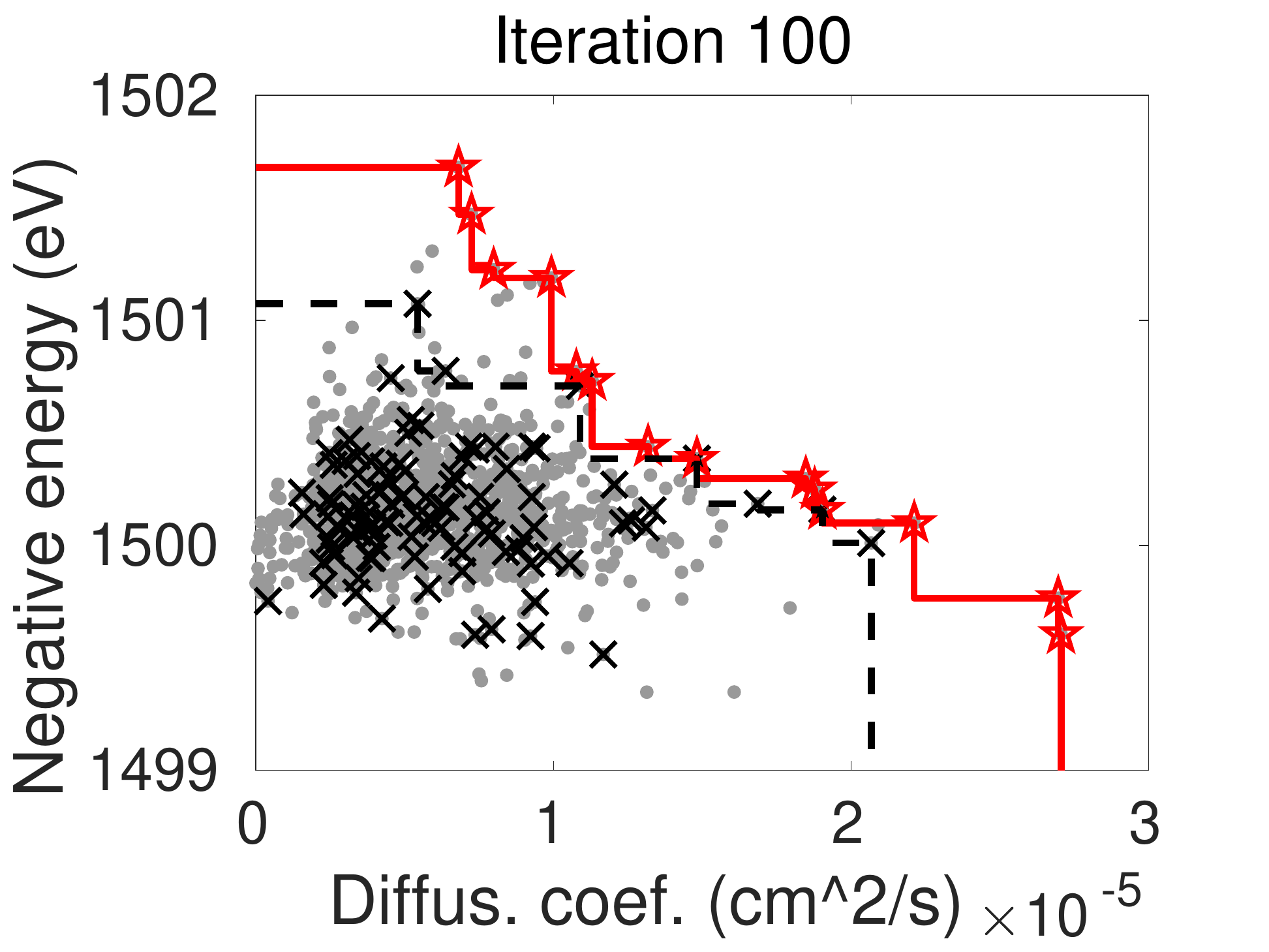} 
  \includegraphics[clip,width=0.24\tw]{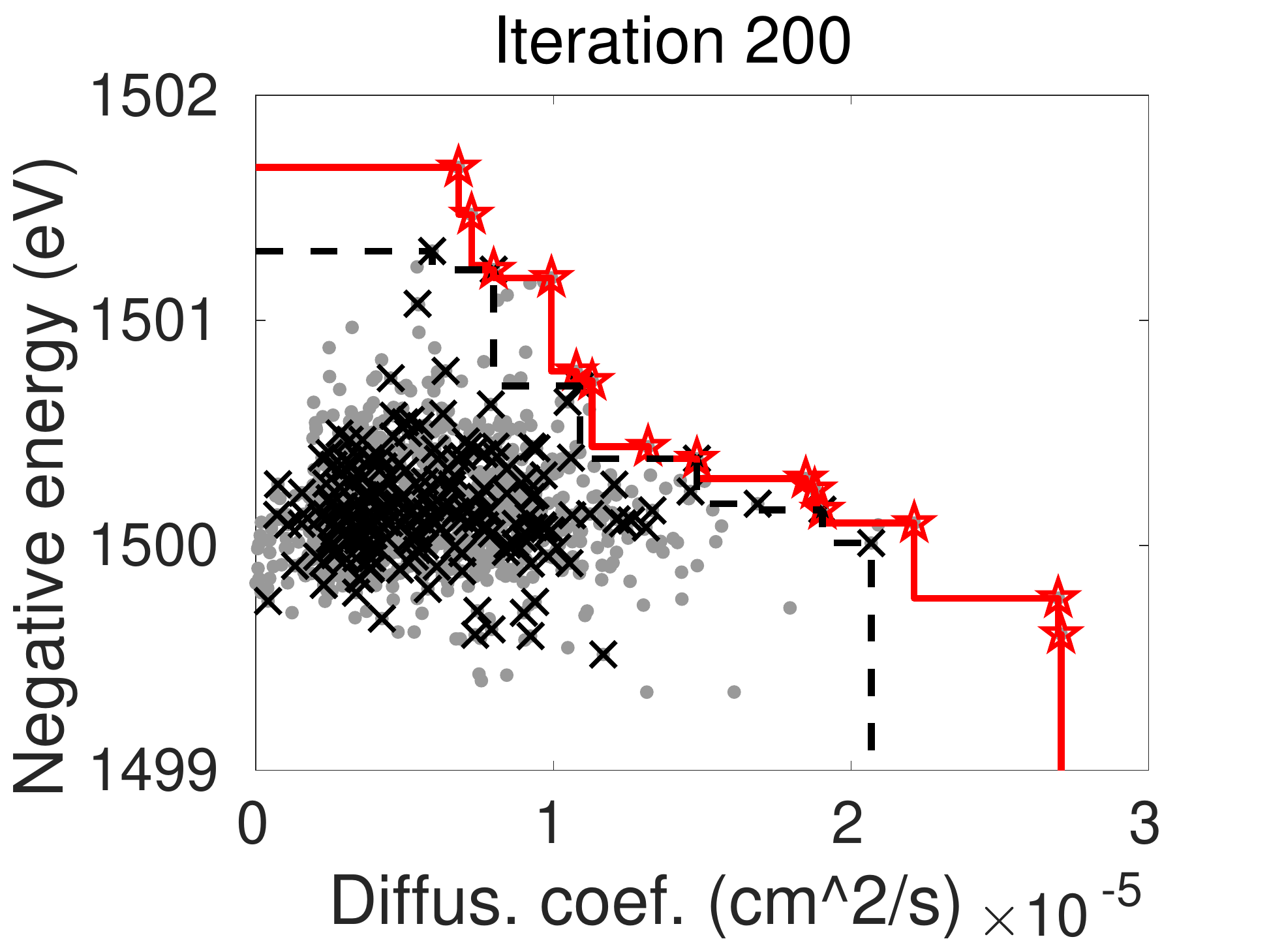} 
  \includegraphics[clip,width=0.24\tw]{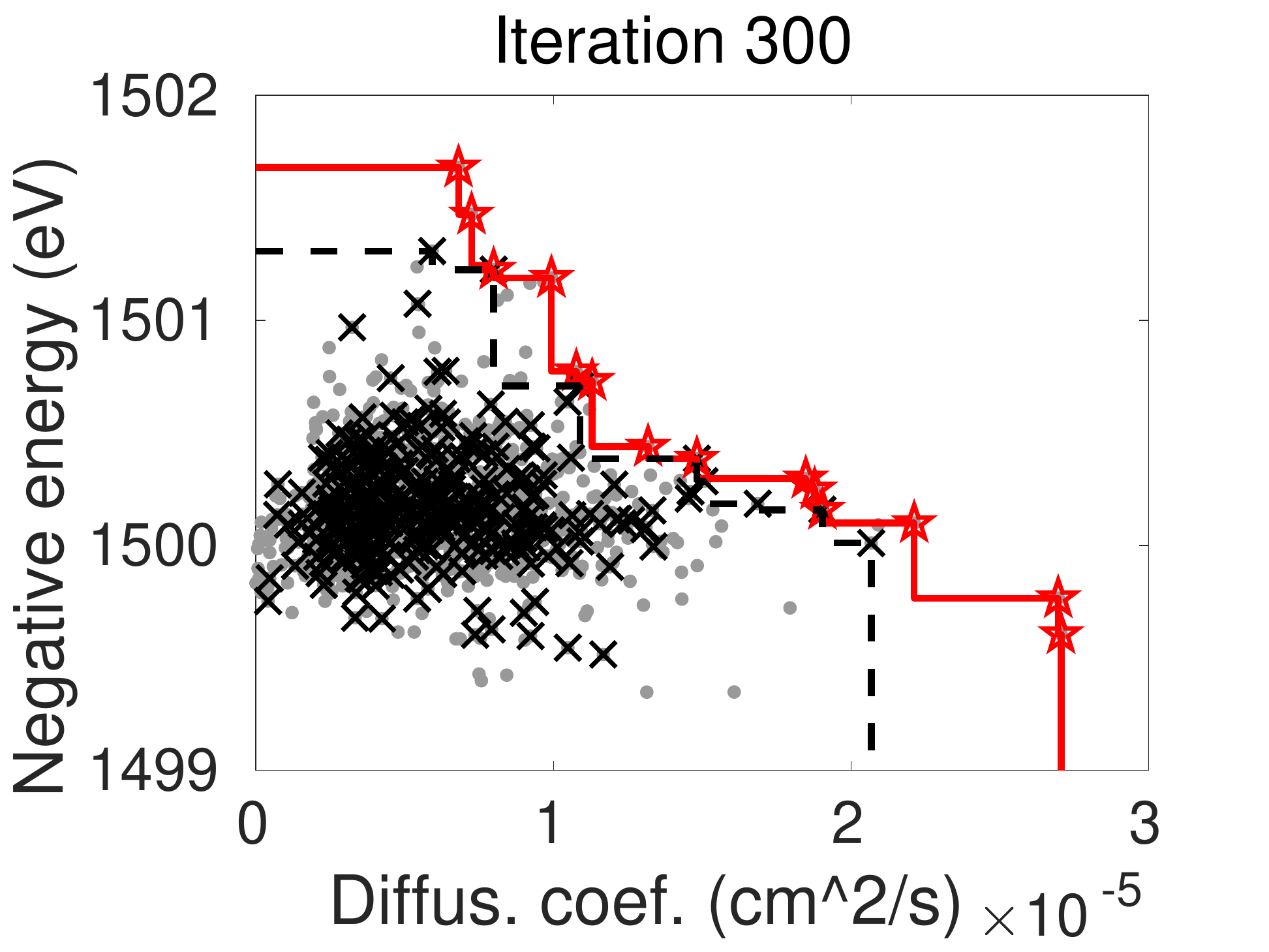} 
  }

 \end{center}
 \caption{
 An example of optimization process for LLTO.
 The vertical axis is the negative value of total energy (eV), and the horizontal axis is diffusion coefficient (cm$^2$/s).
 The red stars are the Pareto optimal points. 
 The black crosses are sampled points for each number of iterations, and the gray dots are candidate points.
 }
 \label{fig:demo-LLTO}
\end{figure}

\begin{figure}[t]
 \begin{center}
  \includegraphics[clip,width=0.4\tw]{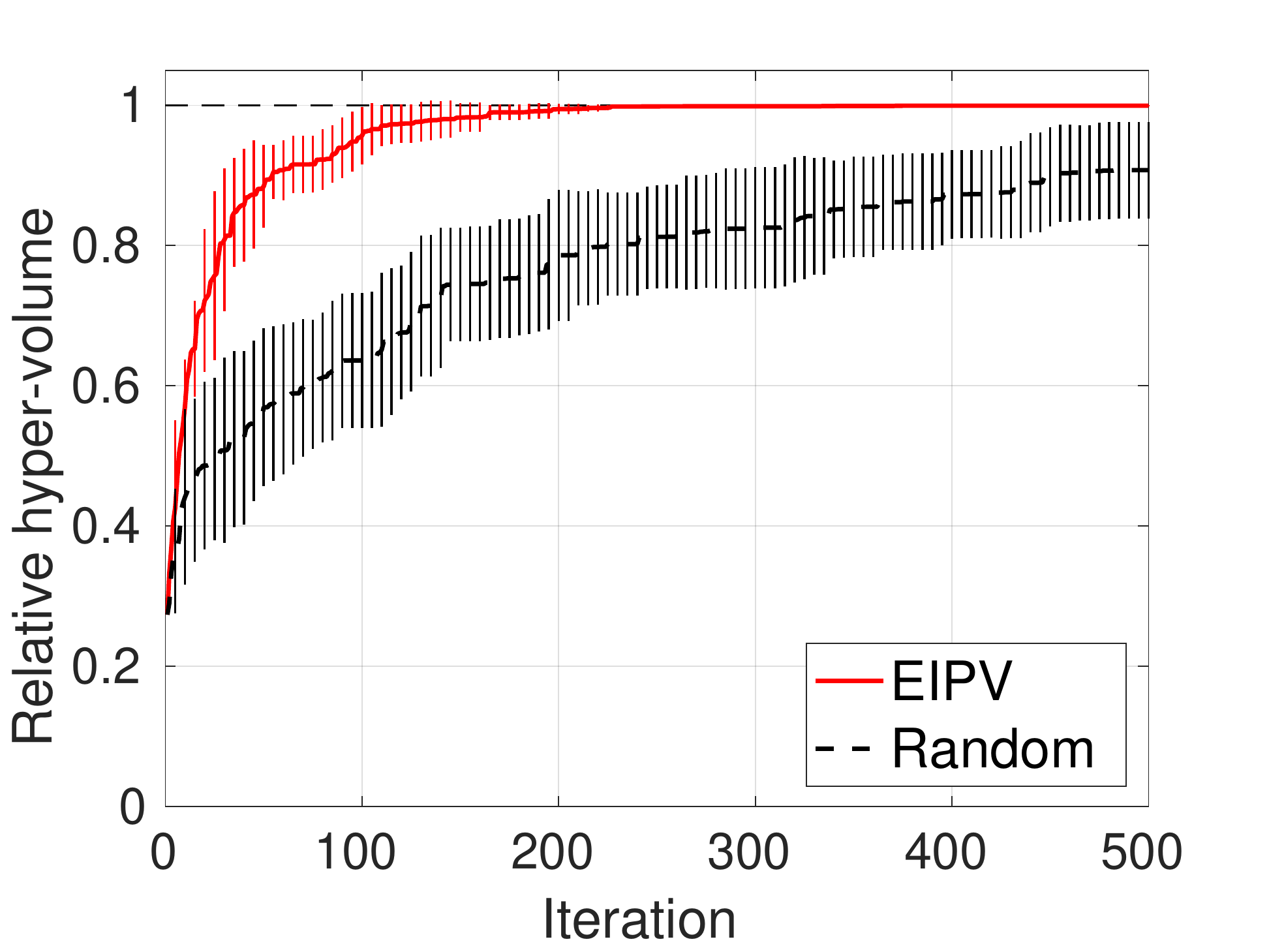} 
 \end{center}
 \caption{
 The transition of the relative hyper-volume for the LLTO data.
 The average of $10$ runs is shown, and the vertical line is standard deviation.
 }
 \label{fig:hvr-LLTO}
\end{figure}

% \figurename~\ref{fig:demo-LLTO} shows (a) the initial state in which $10$ points were randomly selected, (b) the $10$-th iteration, (c) the $50$-th iteration, and (d) the $100$-th iteration.
Figure~\ref{fig:demo-LLTO} shows the Pareto frontier of at the iterations 10, 100, 200, and 300 of EIPV and the random sampling, respectively.
%
% In this case, there are larger number of Pareto optimal points than the case of \ce{Bi2O3}.
%
% Although there exists stronger trade-off relation of conductivity and stability among the candidate points, 
%
In this case, although there are more Pareto optimal points having a stronger trade-off relationship between conductivity and stability when compared with the case of \ce{Bi2O3}, we can see that EIPV more efficiently identified Pareto optimal points than the random sampling.
% we can see that EIPV efficiently identified Pareto optimal points.
%
Figure~\ref{fig:hvr-LLTO} shows the transition of the relative hyper-volume value.
EIPV, once again, rapidly increased the volume substantially faster than the random sampling.

% --------------------------------------------------
% Illustration (LLTO)
% --------------------------------------------------

% --------------------------------------------------
% Hyper-volume (LLTO)
% --------------------------------------------------

% --------------------------------------------------
\section{Discussion} \label{sec:}

In both of the case studies, EIPV achieved a significantly faster optimization than the random sampling.
Unlike the simple single-objective optimization, the multiple optimal points (Pareto set) exist in the multi-objective optimization.
Then, identifying all those optimal points by the random sampling is obviously difficult than the identification a unique optimal point in the single-objective case.
Therefore, building a systematic method of assessment is particularly significant in the ion-conductive material exploration. 
Actually, the concept of Pareto frontier has been used in materials science to evaluate a set of candidate materials in the past \cite{Bligaard2003,Greeley2006,Nunez2018152}.
Another well-known multi-objective exploration approach is the evolutional computation such as the genetic algorithm \cite{Nunez2018152}.
However, this approach often requires a large number of iterations (samplings) because it is based on the random exploration of the candidate space without explicitly modeling the underlying input-output relation.
On the contrary, \cite{Gopakumar2018} applied a machine-learning based multi-objective optimization to a few material search problems, including shape memory alloys.
% , they used different criteria from EIPV. 
%
% They heuristically designed an evaluation criterion of Pareto-frontier, but .
% They heuristically designed an evaluation criterion of Pareto-frontier as $P[I(x)] L$, where $P[I(x)]$ is probability of improvement and $L$ is the ``length'' of the improvement.
% %
% Although they refer to this criterion as the expected improvement, 
% the criterion is not rigorous ``expected value''.
However, they did not show an application of the method to ion-conductive materials.
Further, their criterion is heuristically designed and is difficult to interpret as a direct optimization of an expected value of the Pareto optimality criterion unlike EIPV. 
% without rigorous justification
%
% Although they refer to their criterion as the expected improvement, they did not correctly calculate ``expected value''.
%
% Although their criteria is also applicable to the ion-conductive material search,  
See supplementary appendix~\ref{app:comparison} for further detail, and in our empirical evaluation, EIPV shows a much better performance than that of the above-mentioned method \cite{Gopakumar2018}.
%
% We empirically observed EIPV shows much better performance compared with \cite{} as we show in supplementary appendix~\ref{app:}.

Although we focused on stability and conductivity as the two most fundamental two properties for ion-conductive materials, our framework is general enough to incorporate a variety of properties that can be obtained by using any computational models.
For example, mechanical properties and density can be considered as other important properties for real materials.
EIPV can be directly extended to the case where more than three objective functions exist.
Therefore, simultaneously performing exploration with other such properties would be one of the significant directions for future studies.

\clearpage

% --------------------------------------------------
\section{Conclusions} \label{sec:conclusions}

We proposed a computational framework for efficiently exploring ion-conductive materials that simultaneously exhibit static stability and dynamic conductivity.
Our framework combines theoretical calculations and the Pareto hyper-volume based Bayesian optimization.
We presented two case studies, one on oxygen diffusion in \ce{Bi2O3}, and the other on Li diffusion in La$_{2/3-x}$Li$_{3x}$TiO$_{3}$.
The \ce{Bi2O3} data is evaluated by first-principles molecular dynamics, and the La$_{2/3-x}$Li$_{3x}$TiO$_{3}$ data is evaluated by classical molecular dynamics.
For both of cases, our framework drastically accelerated the exploration of ion-conductive materials as compared with the na{\"i}ve random sampling.

% --------------------------------------------------
\section*{Author Contributions}

{\bf Masayuki Karasuyama}: Conceptualization, Methodology, Software, Visualization, Writing- Original draft preparation.
{\bf Satoru Kimura}: Methodology, Software.
{\bf Tomoyuki Tamura}: Conceptualization, Methodology, Software, Visualization, Writing- Original draft preparation.
{\bf Kazuki Shitara}: Conceptualization, Methodology, Software, Visualization, Writing- Original draft preparation.

% --------------------------------------------------
\section*{Acknowledgment}

This work was supported by ``Materials research by Information Integration'' Initiative (MI$^{2}$I) project of the Support Program for Starting Up Innovation Hub from Japan Science and Technology Agency (JST), MEXT KAKENHI 170H04694, and JST PRESTO JPMJPR15N2.

\clearpage

% --------------------------------------------------
% Reference
% --------------------------------------------------

\bibliographystyle{unsrt}
\bibliography{ref}

\begin{thebibliography}{10}

\bibitem{Seko2015}
A.~Seko, A.~Togo, H.~Hayashi, K.~Tsuda, L.~Chaput, and I.~Tanaka.
\newblock Prediction of low-thermal-conductivity compounds with
  first-principles anharmonic lattice-dynamics calculations and bayesian
  optimization.
\newblock {\em Phys. Rev. Lett.}, 115:205901, Nov 2015.

\bibitem{Toyoura2016}
K.~Toyoura, D.~Hirano, A.~Seko, M.~Shiga, A.~Kuwabara, M.~Karasuyama,
  K.~Shitara, and I.~Takeuchi.
\newblock Machine-learning-based selective sampling procedure for identifying
  the low-energy region in a potential energy surface: A case study on proton
  conduction in oxides.
\newblock {\em Phys. Rev. B}, 93:054112, Feb 2016.

\bibitem{Packwood2017}
D.~Packwood.
\newblock {\em {Bayesian optimization for materials science}}.
\newblock Springer, 2017.

\bibitem{Yonezu2018}
T.~Yonezu, T.~Tamura, I.~Takeuchi, and M.~Karasuyama.
\newblock Knowledge-transfer-based cost-effective search for interface
  structures: A case study on fcc-{Al} [110] tilt grain boundary.
\newblock {\em Phys. Rev. Materials}, 2:113802, 2018.

\bibitem{Emmerich2006}
M.~T.~M. Emmerich, K.~C. Giannakoglou, and B.~Naujoks.
\newblock Single- and multiobjective evolutionary optimization assisted by
  {G}aussian random field metamodels.
\newblock {\em IEEE Transactions on Evolutionary Computation}, 10(4):421--439,
  2006.

\bibitem{Blochl1994}
P.~E. Bl\"ochl.
\newblock Projector augmented-wave method.
\newblock {\em Phys. Rev. B}, 50:17953--17979, 1994.

\bibitem{Kresse1999}
G.~Kresse and D.~Joubert.
\newblock From ultrasoft pseudopotentials to the projector augmented-wave
  method.
\newblock {\em Phys. Rev. B}, 59:1758--1775, 1999.

\bibitem{Kresse1993}
G.~Kresse and J.~Hafner.
\newblock Ab initio molecular dynamics for liquid metals.
\newblock {\em Phys. Rev. B}, 47:558--561, 1993.

\bibitem{Kresse1996}
G.~Kresse and J.~Furthm\"uller.
\newblock Efficient iterative schemes for ab initio total-energy calculations
  using a plane-wave basis set.
\newblock {\em Phys. Rev. B}, 54:11169--11186, 1996.

\bibitem{Perdew1996}
J.~P. Perdew, K.~Burke, and M.~Ernzerhof.
\newblock Generalized gradient approximation made simple.
\newblock {\em Phys. Rev. Lett.}, 77:3865--3868, 1996.

\bibitem{Shitara2017}
K.~Shitara, T.~Moriasa, A.~Sumitani, A.~Seko, H.~Hayashi, Y.~Koyama, R.~Huang,
  D.~Han, H.~Moriwake, and I.~Tanaka.
\newblock First-principles selection of solute elements for er-stabilized bi2o3
  oxide-ion conductor with improved long-term stability at moderate
  temperatures.
\newblock {\em Chemistry of Materials}, 29(8):3763--3768, 2017.

\bibitem{Chen1998}
C.~Chen and J.~Du.
\newblock Lithium ion diffusion mechanism in lithium lanthanum titanate
  solid-state electrolytes from atomistic simulations.
\newblock {\em Journal of the American Ceramic Society}, 98(2):534--542, 2015.

\bibitem{Riquelme2015}
N.~{Riquelme}, C.~{Von L^^c3^^bccken}, and B.~{Baran}.
\newblock Performance metrics in multi-objective optimization.
\newblock In {\em Latin American Computing Conference}.

\bibitem{Rasmussen2006}
C.~E. Rasmussen and C.~K.~I. Williams.
\newblock {\em Gaussian Processes for Machine Learning}.
\newblock MIT Press, Cambridge, MA, USA, 2006.

\bibitem{Shahriari2016}
B.~{Shahriari}, K.~{Swersky}, Z.~{Wang}, R.~P. {Adams}, and N.~{de Freitas}.
\newblock Taking the human out of the loop: A review of {B}ayesian
  optimization.
\newblock {\em Proceedings of the IEEE}, 104(1):148--175, 2016.

\bibitem{Zunger1990}
A.~Zunger, S.-H. Wei, L.~G. Ferreira, and J.~E. Bernard.
\newblock Special quasirandom structures.
\newblock {\em Phys. Rev. Lett.}, 65:353--356, 1990.

\bibitem{Seko2009}
A.~Seko, Y.~Koyama, and I.~Tanaka.
\newblock Cluster expansion method for multicomponent systems based on optimal
  selection of structures for density-functional theory calculations.
\newblock {\em Phys. Rev. B}, 80:165122, 2009.

\bibitem{Seko2010}
A.~Seko.
\newblock Exploring structures and phase relationships of ceramics from first
  principles.
\newblock {\em Journal of the American Ceramic Society}, 93(5):1201--1214,
  2010.

\bibitem{Knauth2002}
P.~Knauth and H.~L. Tuller.
\newblock Solid-state ionics: Roots, status, and future prospects.
\newblock {\em Journal of the American Ceramic Society}, 85(7):1654--1680,
  2002.

\bibitem{Azad1994}
A.~M. Azad, S.~Larose, and S.~A. Akbar.
\newblock Bismuth oxide-based solid electrolytes for fuel cells.
\newblock {\em Journal of Materials Science}, 29(16):4135--4151, 1994.

\bibitem{Sammes1999}
N.~M. Sammes, G.~A. Tompsett, H.~N{\"a}fe, and F.~Aldinger.
\newblock Bismuth based oxide electrolytes― structure and ionic conductivity.
\newblock {\em Journal of the European Ceramic Society}, 19(10):1801 -- 1826,
  1999.

\bibitem{Takahashi1978}
T.~Takahashi and H.~Iwahara.
\newblock Oxide ion conductors based on bismuthsesquioxide.
\newblock {\em Materials Research Bulletin}, 13(12):1447 -- 1453, 1978.

\bibitem{Stramare2003}
S.~Stramare, V.~Thangadurai, and W.~Weppner.
\newblock Lithium lanthanum titanates:~ a review.
\newblock {\em Chemistry of Materials}, 15(21):3974--3990, oct 2003.

\bibitem{Schutt2014}
K.~T. Sch\"utt, H.~Glawe, F.~Brockherde, A.~Sanna, K.~R. M\"uller, and E.~K.~U.
  Gross.
\newblock How to represent crystal structures for machine learning: Towards
  fast prediction of electronic properties.
\newblock {\em Phys. Rev. B}, 89:205118, 2014.

\bibitem{De2016}
S.~De, A.~P Bart{\'o}k, G.~Cs{\'a}nyi, and M.~Ceriotti.
\newblock {Comparing molecules and solids across structural and alchemical
  space}.
\newblock {\em Phys. Chem. Chem. Phys.}, 18(20):13754--13769, 2016.

\bibitem{Ramprasad2017}
R.~Ramprasad, R.~Batra, G.~Pilania, and et~al.
\newblock Machine learning in materials informatics: recent applications and
  prospects.
\newblock {\em npj Comput. Mater.}, 3, 2017.

\bibitem{Tamura2017}
T.~Tamura, M.~Karasuyama, R.~Kobayashi, R.~Arakawa, Y.~Shiihara, and
  I.~Takeuchi.
\newblock Fast and scalable prediction of local energy at grain boundaries:
  machine-learning based modeling of first-principles calculations.
\newblock {\em Modelling and Simulation in Materials Science and Engineering},
  25(7):075003, 2017.

\bibitem{Bligaard2003}
T.~Bligaard, G.~H. J{\`o}hannesson, A.~V. Ruban, H.~L. Skriver, K.~W. Jacobsen,
  and J.~K. N{\o}rskov.
\newblock Pareto-optimal alloys.
\newblock {\em Applied Physics Letters}, 83(22):4527--4529, 2003.

\bibitem{Greeley2006}
J.~Greeley, T.~F. Jaramillo, J.~Bonde, I.~Chorkendorff, and J.~K. N{\o}rskov.
\newblock Computational high-throughput screening of electrocatalytic materials
  for hydrogen evolution.
\newblock {\em Nature Materials}, 5:909 -- 913, 2006.

\bibitem{Nunez2018152}
M.~N{\`u}{\~n}ez-Valdez, Z.~Allahyari, T.~Fan, and A.~R. Oganov.
\newblock Efficient technique for computational design of thermoelectric
  materials.
\newblock {\em Computer Physics Communications}, 222:152 -- 157, 2018.

\bibitem{Gopakumar2018}
A.~M. Gopakumar, P.~V. Balachandran, and D.~et~al. Xue.
\newblock Multi-objective optimization for materials discovery via adaptive
  design.
\newblock {\em Sci. Rep.}, 8:3738, 2018.

\end{thebibliography}

\clearpage

% --------------------------------------------------
% Appendix
% --------------------------------------------------

\appendix

\counterwithin{figure}{section}
\renewcommand{\thetable}{\Alph{section}.\arabic{table}}
\renewcommand{\thefigure}{\Alph{section}.\arabic{figure}}

\section{Comparison with EI-Centroid and EI-Maximin}
\label{app:comparison}

The probability of improvement for the multi-objective optimization is defined as 
\begin{align*}
 p(I(\*x)) =
 \int_{\*y \in \cY_{\rm ND}}
 \phi_{\*x}(\*y) 
 % p(\*y \mid \*x) 
 \mathrm{d} \*y,
\end{align*}
where $\cY_{\rm ND}$ is the output space that is not dominated by current observed points, and 
% $p(\*y \mid \*x)$ 
$\phi_{\*x}(\*y)$ 
is the Gaussian predictive distribution of the machine-learning model.
In \cite{Gopakumar2018}, the acquisition function is defined as $p(I(\*x)) L$, where $L$ is a ``length'' parameter that evaluates the degree of improvement.
% , they further introduced the ``length'' parameter $L$.
%
The following two approaches, called (a) EI-Centroid, and (b) EI-Maximin, are shown to define $L$ in \cite{Gopakumar2018}.
\begin{itemize}
 \item[(a)] The EI-Centroid approach is defined as the distance between centroid of the predictive distribution and the closest point on the Pareto front. 
	    The centroid $(c_1(\*x), c_2(\*x))$ is first fixed as an expected value of the predictions in $\cY_{\rm ND}$, and then, $L$ is calculated as 
	    $L = \sqrt{ (c_1(\*x) - \tilde{y}_1(\*x))^2 + (c_2(\*x) - \tilde{y}_2(\*x)) }$, 
	    where 
	    $(\tilde{y}_1(\*x), \tilde{y}_2(\*x))$ is the closest point on the Pareto front to the centroid.
	    However, this $L$ is not an expected value of the improvement, because the centroid and its closest point on the Pareto front is fixed in the calculation of $L$. 
	    %
	    % This $L$ should be referred as an improvement by centroid.
	    %
	    To define the expected improvement in terms of this distance, the expectation over the entire distance function should be taken as
	    $\int_{\*y \in \cY_{ND}} \sqrt{ (y_1(\*x) - \tilde{y}_1(\*x))^2 + (y_2(\*x) - \tilde{y}_2(\*x)) } \phi_{\*x}(\*y) / P(I(\*x)) \mathrm{d} \*y$.
	    %, where $(y_1(\*x),y_2(\*x))$ is a random prediction by a machine-learning model, and 
	    % $(y_1(\*x), y_2(\*x))$ is the closed point on the Pareto front from $(f_1(\*x),f_2(\*x))$.
	    Note that, here, 
	    $(\tilde{y}_1(\*x), \tilde{y}_2(\*x))$ must be defined for each one of $(y_1(\*x),y_2(\*x))$ in this integral, but this computation is complicated to perform.
	    % To define the expected improvement of this distance, expectation over the entire
	    % $\sqrt{ (F_1(\*x) - f_1(\*x))^2 + (F_1(\*x) - f_2(\*x)) }$
	    % should have been taken, where $(f_1(\*x),f_2(\*x))$ is a random prediction by a machine-learning model, and $(F_1(\*x), F_2(\*x))$ is the closed point on the Pareto front from $(f_1(\*x),f_2(\*x))$.

 \item[(b)] The EI-Maximin approach is defined as the distance
	    % $L = \max_i (\min(p_{i1} - \mu_1(\*x), p_{i2} - \mu_2(\*x)), 0)$, 
	    $L = \max_i (\min(\mu_1(\*x) - p_{i1}, \mu_2(\*x) - p_{i2}), 0)$, 
	    where 
	    $(p_{i1}, p_{i2})$
	    is a point in the current Pareto optimal, and 
	    $(\mu_1(\*x),\mu_2(\*x))$
	    is the predictive mean
	    (Note that we slightly modified $L$ so that it can be adopted to the maximization problem while the original publication \cite{Gopakumar2018} formulates the problem as the minimization).     
	    This $L$ is also not an expected improvement, because $(\mu_1(\*x), \mu_2(\*x))$ is fixed inside $L$.
	    Here, again, to define the expected improvement, the expectation over the entire distance function
	    % $\int_{\*y \in \cY_{ND}} \max_i (\min(p_{i1} - y_1(\*x), p_{i2} - y_2(\*x))) \phi_{\*x}(\*y) / P(I(\*x)) \mathrm{d} \*y$
	    $\int_{\*y \in \cY_{ND}} \max_i (\min(y_1(\*x) - p_{i1} ,  y_2(\*x)- p_{i2}), 0) \phi_{\*x}(\*y) / P(I(\*x)) \mathrm{d} \*y$
	    should be considered, but this is also computationally difficult in practice.

\end{itemize}
%
% In this sense, these criteria do not correspond to any ``expected value'' of the improvement.
In both cases, the expectation is taken ``inside'' of $L$, because of which it is difficult to interpret the quantity $P(I(\*x)) L$ as an expectation of an evaluation measure of the Pareto optimality.
%
% Actually, each of them is the improvement by one fixed prediction (centroid or predictive mean)
%
% EIPV takes the expectation of the increase of Pareto hyper-volume 
% while the expectation of improvement should be taken over ``outside'' of the evaluation measure as we show above.
%
In contrast, EIPV directly evaluates the expected value of the increase of the Pareto hyper-volume, which is a standard evaluation measure of optimality of Pareto optimization as we mentioned in the main text. %  (expectation is taken over ``outside'' of the increase of the volume as shown in the end of Section~\ref{ssec:EIPV}).

\begin{figure}[t]
 \begin{center}
  \includegraphics[clip,width=0.4\tw]{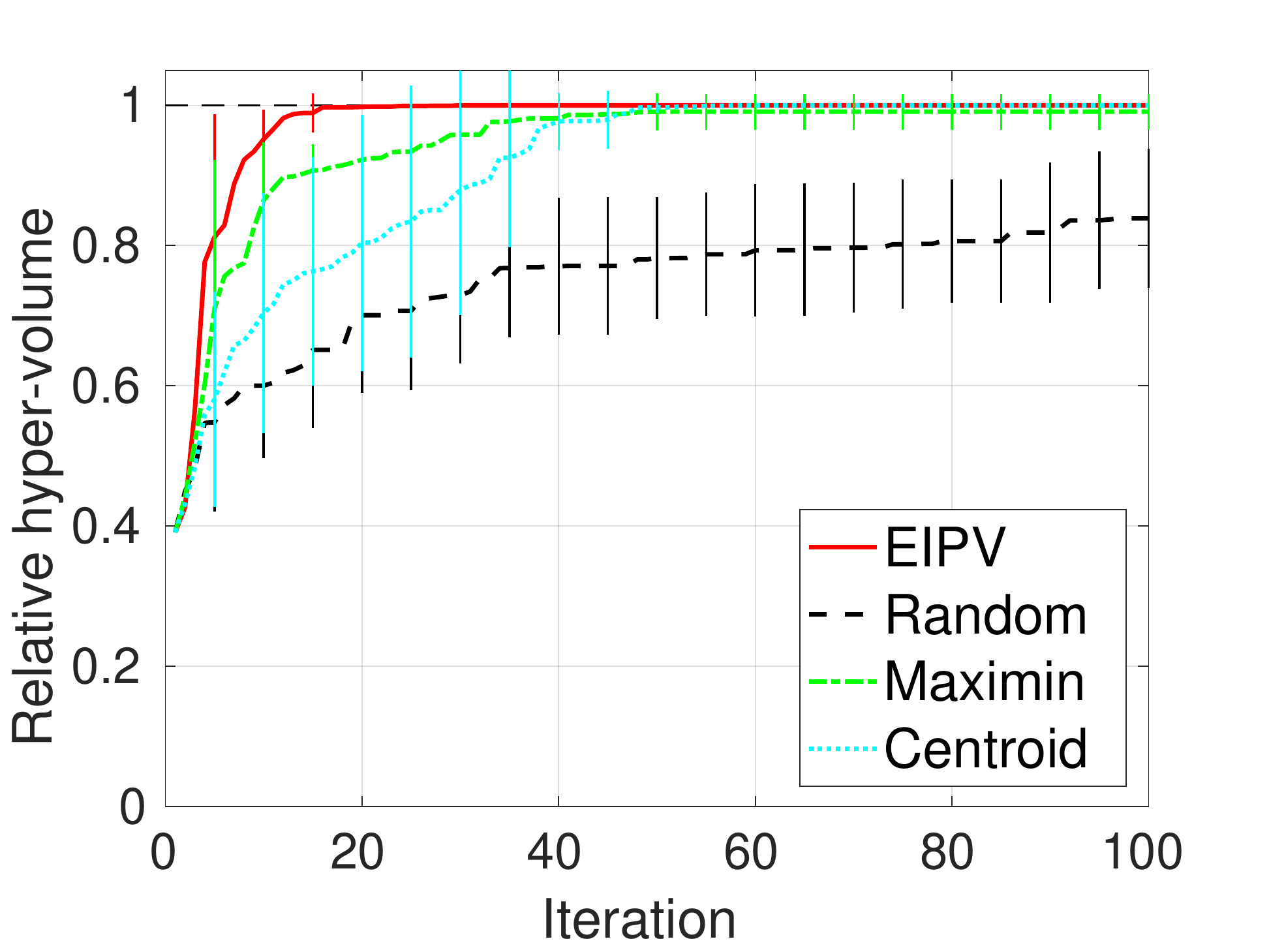} 
 \end{center}
 \caption{ 
 The transition of the relative hyper-volume for the \ce{Bi2O3} data.
 %
 % The average of $10$ runs is shown, and the vertical line is standard deviation.
 }
 \label{fig:Bi2O3-compare}
\end{figure}

\begin{figure}[t]
 \begin{center}
  \includegraphics[clip,width=0.4\tw]{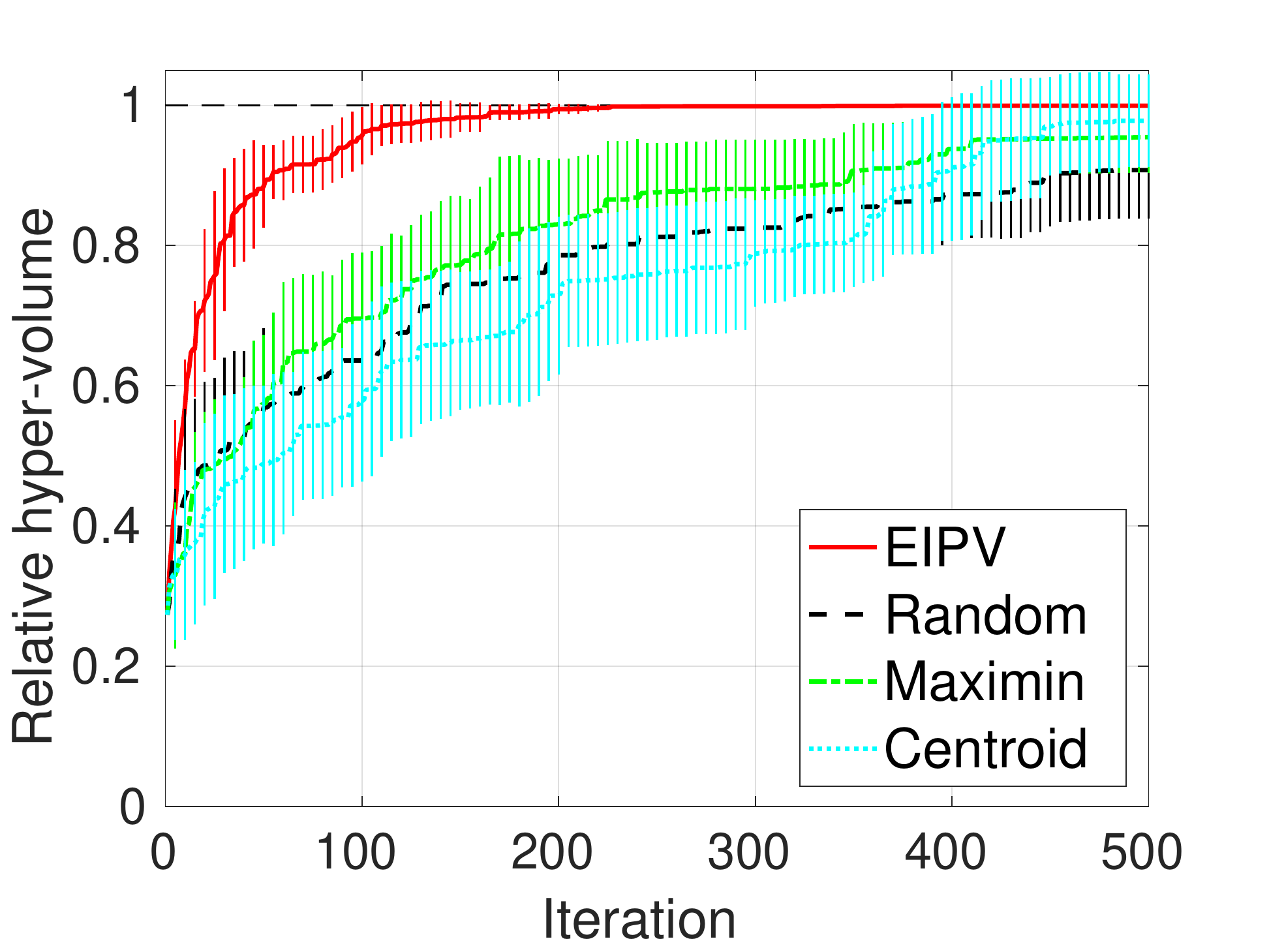} 
 \end{center}
 \caption{
 The transition of the relative hyper-volume for the LLTO data.
 %
 % The average of $10$ runs is shown, and the vertical line is standard deviation.
 }
 \label{fig:LLTO-compare}
\end{figure}

Figure~\ref{fig:Bi2O3-compare} and \ref{fig:LLTO-compare} show comparison of EIPV with EI-Centroid (Centroid) and EI-Maximin (Maximin).
The setting is the same as in Section~\ref{sec:result}.
From the definition, $L$ of Maximin could be $0$ for all the candidates.
When it happened, we used the Centroid criterion at that iteration.
We obviously see that EIPV is significantly more efficient than the Centroid and Maximin for both the cases.

% --------------------------------------------------
% Hyper-volume (Bi2O3) with four methods
% --------------------------------------------------

% --------------------------------------------------
% Hyper-volume (LLTO) with four methods
% --------------------------------------------------

\end{document}